\documentclass[11pt,letter]{article}

\usepackage{graphicx}
\usepackage{amsmath}
\usepackage{amssymb}
\usepackage{theorem}
\usepackage{subfigure}
\usepackage{multirow}

\theoremstyle{plain}
\theorembodyfont{\rm}
\newtheorem{thm}{Theorem}
\newtheorem{property}{Property}

\newtheorem{lma}{Lemma}

\newtheorem{remark}{Remark}
\newtheorem{conjecture}{Conjecture}
\newtheorem{pf}{Proof}

\newcommand{\argum}{\mathit{arg}}

\newcommand{\Real}{\mathbb{R}}
\newcommand{\Natural}{\mathbb{N}}

\newcommand{\Omit}[1]{}

\begin{document}

\vspace*{10mm}
\begin{center}
\begin{minipage}{170mm}
\begin{center}
{\LARGE The Gathering Problem for Two Oblivious Robots \\
        with Unreliable Compasses}\footnotemark[1]
\end{center} 
\end{minipage} \\
\vspace*{10mm}
Taisuke Izumi\footnotemark[2]$^{,}$\footnotemark[5] \hspace{3mm} 
Samia Souissi\footnotemark[2] \hspace{3mm} 
Yoshiaki Katayama\footnotemark[2] \hspace{3mm} 
Nobuhiro Inuzuka\footnotemark[2] \hspace{3mm} \\
Xavier D\'{e}fago\footnotemark[3] \hspace{3mm} 
Koichi Wada\footnotemark[2] \hspace{3mm}
Masafumi Yamashita\footnotemark[4] 
\footnotetext[1]{Preliminary versions of this paper appeared as: 
``Gathering autonomous mobile robots with dynamic compasses: 
An optimal result,'' 
{\em Proc. International Symposium on Distributed Computing,}
298--312, 2007,         
``Gathering two stateless mobile robots using very inaccurate 
compasses in finite time,'' 
{\em Proc. International Conference on Robot 
Communication and Coordination,} 48, 2007,
and ``Gathering problem of two asynchronous mobile robots 
with semi-dynamic compasses,'' 
{\em Proc. International Colloquium on Structural 
Information and Communication Complexity,} 5--19, 2008.}
\footnotetext[2]{Graduate School of Engineering, 
Nagoya Institute of Technology, Gokiso-cho, 
Showa-ku, Nagoya, Aichi, 466-8555, Japan.} 
\footnotetext[3]{Graduate School of Information Science,
Japan Advanced Institute of Science and Technology (JAIST)
1-1 Asahidai, Tatsunokuchi, Nomigun Ishikawa 923-1292, Japan.}
\footnotetext[4]{Department of Informatics,
Kyushu University 744, Motooka, Fukuoka, 819-0395, Japan.}
\footnotetext[5]{Corresponding Author. E-mail: t-izumi@nitech.ac.jp. 
Tel:+81-52-735-5567, Fax:+81-52-735-5408}
\vspace*{13mm} \\
{\bf Abstract} \\
\vspace*{7mm}
\begin{minipage}{170mm}

Anonymous mobile robots are often classified into synchronous, 
semi-synchronous and asynchronous robots 
when discussing the pattern formation problem.
For semi-synchronous robots,
all patterns formable with memory are also formable {\em without} memory, 
with the single exception of forming a point (i.e., the gathering) by two robots.
(All patterns formable with memory are formable without memory 
for synchronous robots, and little is known for asynchronous robots.)
However, the gathering problem for two semi-synchronous robots without memory
(called oblivious robots in this paper) is trivially solvable 
when their local coordinate systems are consistent,
and the impossibility proof essentially uses the inconsistencies 
in their coordinate systems.
Motivated by this,
this paper investigates the magnitude of consistency 
between the local coordinate systems necessary and sufficient 
to solve the gathering problem for two oblivious robots under
semi-synchronous and asynchronous models.
To discuss the magnitude of consistency,
we assume that each robot is equipped with an {\em unreliable} compass,
the bearings of which may deviate from an absolute reference direction,
and that the local coordinate system of each robot is determined by its compass.
We consider two families of 
unreliable compasses, namely,
\emph{static compasses} with (possibly incorrect) constant bearings,
and \emph{dynamic compasses} the bearings of which can change arbitrarily
(immediately before a new look-compute-move cycle starts and
after the last cycle ends).

For each of the combinations of robot and compass models,
we establish the condition on deviation $\phi$ 
that allows an algorithm to solve the gathering problem, 
where the deviation is measured by the largest angle formed 
between the $x$-axis of a compass and the reference direction 
of the global coordinate system:
$\phi < \pi/2$ for semi-synchronous and asynchronous robots with static compasses, 
$\phi < \pi/4$ for semi-synchronous robots with dynamic compasses, 
and $\phi < \pi/6$ for asynchronous robots with dynamic-compasses. 
Except for asynchronous robots with dynamic compasses, 
these sufficient conditions are also necessary.

\end{minipage} \\
\end{center}

\section{Introduction} \label{secIntroduction}

Geometric pattern formation by anonymous mobile robots 
have gained much attention 
\cite{AP06,AOSY99,BPT09-1,CFPS03,CP02,CP05,CP08,FPSW05,KTIIW07,P05,SDY05,SY99,YIKIW09}.
In the literature, a robot is represented by a point 
and repeatedly executes a ``look-compute-move'' cycle, during which, 
it observes the positions of all robots (\textsc{look} phase), 
computes the next position using a given algorithm (\textsc{compute} phase), 
and moves to that position (\textsc{move} phase).  
A robot does not have access to a global coordinate system,
and all its computations are done in terms of its local coordinate system.
The robots do not have identifiers, 
are not equipped with communication devices,
and execute the same algorithm.

The robots' behaviors are in general asynchronous.
Their executions of the look, compute and move phases may be interleaved
in the sense that a robot may observe, for instance, another robot 
while it is moving.\footnote{
The robot however cannot determine its velocity,
in particular, whether or not it is moving.}
The robots are said to be \emph{semi-synchronous} when the execution of 
their cycles is assumed to be ``instantaneous,''
which intuitively means that a robot is never observed 
while it is moving.
Robots are said to be \emph{synchronous},
if all of them always execute the instantaneous cycles simultaneously.
A robot is said to be \emph{oblivious},
if it has no memory to remember its execution history, 
and its computations depend only on what it is observing in the current cycle. 
A robot is said to be \emph{non-oblivious},
if it has sufficient memory to remember the whole execution history
and its action can depend also on what it has observed in the past.

The set of patterns formable by semi-synchronous oblivious robots is,
by definition, a subset of the patterns formable 
by semi-synchronous non-oblivious robots.
This inclusion relation is proper since
the point formation (i.e., the gathering) problem for two robots 
is solvable for semi-synchronous non-oblivious robots,
but it is unsolvable for semi-synchronous oblivious robots,
which exhibits the impact of memory in forming a pattern~\cite{SY99}.
Note that the gathering problem for more than two
semi-synchronous oblivious robots is solvable
provided that a robot can count the number of robots residing at the same 
point (i.e., detect multiple robots residing at the same point).
Interestingly, with the sole exception of gathering of two robots mentioned above, any pattern formable by semi-synchronous non-oblivious robots 
is also formable by semi-synchronous oblivious robots~\cite{YS09}.
Thus, the memory helps only in the case of gathering two robots.
All patterns formable by non-oblivious synchronous robots 
are formable by oblivious synchronous robots, and little is 
known for asynchronous robots.
These facts motivate our study of the gathering problem 
for two oblivious semi-synchronous and asynchronous robots.

The impossibility proof of the gathering problem for
two oblivious semi-synchronous robots relies on the ``full'' inconsistency 
of their local coordinate systems \cite{SY99},
while there is a simple gathering algorithm 
when they are ``fully'' consistent.
A natural question then arises;
what is the minimum magnitude of consistency between the local coordinate systems 
that is necessary and sufficient to solve the gathering problem for two 
oblivious robots?
We answer this question in the paper
for both semi-synchronous and asynchronous robots.

To discuss the magnitude of consistency,
we consider that a robot is equipped with an {\em unreliable} compass,
the bearings of which may deviate from the absolute ones 
(i.e., the bearings of global coordinate system),
and assume that the compass determines the local coordinate system.

We consider two families of unreliable compasses with respect to
the difference of timings that a compass can change the bearings.
A {\em static compass} never changes its (possibly incorrect) bearings 
once an execution of algorithm starts.
A {\em dynamic compass}, on the other hand, 
can change the bearings arbitrary times immediately before
a new look-compute-move cycle starts,
after the last cycle ends. 
We can consider a more general family of compasses
which can change the bearings even during the execution 
of a look-compute-move cycle.
We however do not investigate this case in this paper,
since the impossibility of gathering in this case is trivial.

To measure the magnitude of deviation of a compass from 
the global coordinate system, 
we use the angle formed by the $x$-axis of the compass 
and the reference direction of the global coordinate system.
In this paper,
we investigate the maximum deviation that is necessary and sufficient 
for two oblivious robots to solve the gathering problem.
We consider each of the four combinations of robot and compass models,
and essentially show the following results:

\begin{description}
\item[Semi-Synchronous Robots with Static Compasses (SS):]
There is a gathering algorithm for two oblivious semi-synchronous robots
that uses static compasses with maximum deviation $\phi$,
if and only if $0 \leq \phi < \pi/2$.

\item[Semi-Synchronous Robots with Dynamic Compasses (SD):]
There is a gathering algorithm for two oblivious semi-synchronous robots
that uses dynamic compasses with maximum deviation $\phi$,
if and only if  $0 \leq \phi < \pi/4$.

\item[Asynchronous Robots with Static Compasses (AS):]
There is a gathering algorithm for two oblivious asynchronous robots
that uses static compasses with maximum deviation $\phi$,
if and only if $0 \leq \phi < \pi/2$.

\item[Asynchronous Robots with Dynamic Compasses (AD):]
There is a gathering algorithm for two oblivious asynchronous robots
that uses dynamic compasses with maximum deviation $\phi$,
if $0 \leq \phi < \pi/6$.
\end{description}

Note that whether or not $0 \leq \phi < \pi/6$ is necessary 
is left as an open problem for asynchronous robots with dynamic
compasses. 

The remainder of this paper is organized as follows:
After briefly surveying related works in Section~\ref{Srelatedworks}, 
Section \ref{secPreliminaries} defines formal models of robots and compasses.
We discuss the solvability of Gathering by semi-synchronous and 
asynchronous robots with compasses
in Sections~\ref{Ssemisynch} and \ref{Sasynch}, respectively.
Finally, Section~\ref{Sconc} concludes the paper.

\section{Related Works}
\label{Srelatedworks}

The set of geometric patterns formable/convergable%
\footnote{\emph{Formation} requires that all robots form 
the pattern within a finite number of steps, while 
\emph{convergence} only requires the robots to approach the pattern asymptotically.}
by a set of anonymous semi-synchronous robots was characterized 
by Suzuki and Yamashita for non-oblivious robots \cite{SY99}
and also for oblivious robots \cite{YS09}. 
From these two studies, it turns out that memory can help with 
the formation/convergence 
of geometric patterns only in very specific cases.
Indeed, non-oblivious and oblivious semi-synchronous robots 
can solve formation/convergence for the same set of geometric patterns, 
except for the formation of a point with exactly two robots\footnote{
	Two oblivious robots can converge to a point with a naive algorithm
	that consists of always moving toward the midpoint
        of their positions.}
(i.e. the gathering of two robots).
As for asynchronous robots, little is known,
except that the gathering problem for more than 
two robots is solvable~\cite{CFPS03,CP02}.
These positive results for the gathering problem rely on the ability of
robots to detect multiplicity or, 
in other words, the ability to count the number of robots 
that share a given location. 
This assumption is indeed necessary. Otherwise, 
the gathering for more than two robots is reducible 
to the problem of two robots~\cite{P05}.

Essentially, the difficulty in forming (and even in converging to) 
a pattern by robots lies in the difficulty of breaking
symmetry among the robots.
In fact, any pattern is formable,
given a symmetry-breaking tool, like a compass.
The use of a compass was first introduced by Flocchini \mbox{et al.}~\cite{FPSW05}.
They showed that asynchronous robots with limited visibility can solve 
the gathering problem when every robot has access to a \emph{correct} compass.
Souissi \mbox{et al.}~\cite{SDY05} extended the above result 
to the situation where compasses are \emph{eventually consistent}.
A compass is said to be eventually consistent 
if it is unstable and inaccurate for some arbitrary long period,
but eventually stabilizes to show the accurate direction.

In contrast, an {\em unreliable} compass may never correctly indicate 
the correct direction,
although its maximum deviation is bounded.
This type of compass was first introduced by Katayama \mbox{et al.}~\cite{KTIIW07}.
They showed that the gathering problem for two oblivious asynchronous 
robots is solvable if their compasses are either
1) static whose deviation is less than $\pi/6$, or
2) dynamic whose deviation is less than $\pi/8$.

Some other work has focused on fault-tolerant formation/convergence 
for anonymous robots.
Let $F$ be the number of faulty robots.
Cohen \mbox{et al.}~\cite{CP05} showed that convergence 
to a point is solvable for $n$ asynchronous robots by simply
converging to their center of gravity,
even if some of the robots may possibly crash 
(as long as there exists a non-faulty robot).
Bouzid \mbox{et al.}~\cite{BPT09-1,BPT09-2} proposed three Byzantine 
resilient convergence algorithms in one-dimensional space;
1) for synchronous robots provided $n>2F$,
2) for semi-synchronous robots provided $n>3F$, and
3) for asynchronous robots provided $n>4F$.
Agmon \mbox{et al.}~\cite{AP06} showed that 
1) there is no Byzantine resilient gathering algorithm 
for semi-synchronous robots even if $F = 1$, and
2) there is a Byzantine resilient gathering algorithm 
for synchronous robots if and only if $n \geq 3F+1$.

Finally, effects of sensor/control errors in convergence to 
a point were discussed by Cohen and Peleg~\cite{CP08}
and Yamamoto \mbox{et al.}~\cite{YIKIW09},
assuming that the robots are aware of the global coordinate system.
They measured sensor/control errors by a pair of the maximum 
angle and distance errors, 
and obtained necessary and/or sufficient conditions for robots 
to have a convergence algorithm in terms of the pair.

\section{System Model and Problem Definition} 
\label{secPreliminaries}

\noindent
{\bf Robot with Compass:} ~~

In this paper,
we investigate an autonomous mobile robot system $\cal R$ 
consisting of two oblivious robots $r_0$ and $r_1$ working 
in a two dimensional Euclidean space $\Real^2$,
where $\Real$ is the set of real numbers.
The robots are {\em anonymous} and do not have identifiers;
the subscript~$i$ of $r_i$ is used only for the purpose of explanation.
Let $\mathbf{r}_i(t)$ be the coordinates in the global $x$-$y$ 
coordinate system~$Z$ of a robot~$r_i$ at time~$t$.
The {\em configuration}~$C(t)$ of $\cal R$ 
at time~$t$ is defined by $(\mathbf{r}_0(t), \mathbf{r}_1(t))$.

The robots do not have access to $Z$,
and each robot~$r_i$ at time~$t$ observes, computes and moves
in its local $x$-$y$ coordinate system $Z_{(i,t)}$.
The origin of $Z_{(i,t)}$ is always at the current position of $r_i$, 
and the direction of the $x$-axis corresponds to the bearings of its compass.
$Z$, $Z_{(0,t)}$ and $Z_{(1,t)}$ are right-hand systems.
Thus, for any point with coordinates $\mathbf{p}$ in $Z$,
its coordinates $Z_{(i,t)}(\mathbf{p})$ in $Z_{(i,t)}$ are calculated by:
\[
Z_{(i,t)}(\mathbf{p})^T = sc_i(t) \left(
\begin{array}{rr}
\cos \phi_i(t) & \sin \phi_i(t)\\
- \sin \phi_i(t) & \cos \phi_i(t) \\ 
\end{array}
\right) (\mathbf{p} - \mathbf{r}_i(t))^T,
\]
where the {\em scaling ratio} $sc_i(t)$ (with $0 < sc_i(t) < \infty$)
is the ratio of the unit length in $Z$ to that in $Z_{(i,t)}$,
and the {\em deviation} $\phi_i (t)$ (with $-\pi < \phi_i (t) \leq \pi$)
is the angle formed by the $x$-axes of $Z$ and $Z_{(i,t)}$.
The deviation abstracts the compass,
and $Z_{(i,t)}(\mathbf{r}_i(t)) = \mathbf{0}$ always holds.
Since the scaling ratio and the compass (i.e., deviation) may change 
as time goes,
the local coordinate system $Z_{(i,t)}$ may change accordingly.

\medskip
\noindent
{\bf Look-Compute-Move Cycle of Robot:}~~

Each oblivious robot $r_i$ repeatedly executes the look-compute-move cycle.
The local coordinate system $Z_{(i,t)}$, 
and thus both the scaling ratio $sc_i(t)$ 
and the compass deviation $\phi_i (t)$ remain unchanged 
during a cycle (i.e., from \textsc{look} to \textsc{move}).

Suppose that a robot, say $r_0$, starts executing the cycle at time~$t_0$.
In the look phase, $r_0$ observes the other robot~$r_1$ and obtains
the coordinates of $r_1$'s position
in the local coordinate system $Z_{(0,t_0)}$.
We assume that this observation is an instantaneous action;
$r_0$ obtains $Z_{(0,t_0)}(\mathbf{r}_1(t))$ as the result of observation,
where $t$ is a time instant in the look phase.
If the gathering has already been achieved, then
$r_0$ observes exactly one point at the origin.\footnote{
The converse (one point at origin implies gathering) 
may not hold for asynchronous robots. 
See the definition of the gathering problem.}

Next, robot~$r_0$ computes, based on the coordinates 
$Z_{(0,t_0)}(\mathbf{r}_1(t))$,
the coordinates in $Z_{(0,t_0)}$ of its next position. 
The algorithm is simply a total function $\psi$ on $\Real^2$.
That is, when the compute phase finishes, 
$r_0$ obtains $\psi(Z_{(0,t_0)}(\mathbf{r}_1(t)))$, 
as the coordinates of its next position.

In the move phase, 
$r_0$ moves linearly toward coordinates 
$\psi(Z_{(0,t_0)}(\mathbf{r}_1(t)))$ in $Z_{(0,t_0)}$
at a (possibly variable) finite speed that $r_0$ cannot control.
Since $Z_{(i,t)}$ does not change during the look-compute-move cycle,
$Z_{(0,t_0)}$ is the current local coordinate system of $r_0$.
The move phase may be too short for $r_0$ to reach the next position,
and thus $r_0$ may finish the current execution of the look-compute-move 
cycle on the way to its next position.
We however assume that the move phase is long enough to move over 
a small distance~$\delta$ (in $Z$).\footnote{
Obviously, no gathering algorithm exists if a robot
can finish the move phase at any position between the current and the
next position.}

We make three simplifying assumptions that incur no loss of generality.
Let $\Natural$ denote the set of non-negative integers.
\begin{enumerate}
\item
Each execution of the look-compute-move cycle starts at the time 
at which the observation action is taken in the cycle.
\item
The system is initialized at time~0,
i.e., the first observation action is taken by a robot at $0$.
\item
The set of time instants at which the robots start executions
of the look-compute-move cycle 
(or equivalently, the time instants at which they take observation 
actions) is $\Natural$.
A robot is said to be {\em activated} at  a time~$t \in \Natural$, 
if it starts executing the cycle at $t$.
\end{enumerate}

\medskip
\noindent
{\bf Execution:}~~

Given an algorithm and an initial configuration~$C(0)$,
let us observe the behavior of robot system~$\cal R$.
Let $C(t)$ be the configuration of $\cal R$ at $t \in \Natural$.
An infinite sequence~${\cal E}: C(0), C(1), \ldots$ is
called an {\em execution} of~$\cal R$.
Recall that $C(t)$ is the configuration at time $t$,
in which at least one robot is activated.
An execution must be {\em fair} in the sense that
both robots are activated infinitely many times in any infinite execution.

\medskip
\noindent
{\bf Asynchronous and Semi-Synchronous Robots:}~~

Robots are said to be {\em asynchronous} if we do not make
any assumption on the execution of the look-compute-move cycle.
Thus, a robot may be moving (move phase)
while the other robot starts the look phase.

Robots are said to be {\em semi-synchronous} 
if every execution of the cycle is instantaneous.
An execution of the cycle started at time~$t \in \Natural$ is 
said to be {\em instantaneous},
if the look and the compute phases immediately finish at $t$
and the move phase finishes before $t+1$.

\medskip
\noindent
{\bf Static and Dynamic Compasses:}~~

A compass is a $\phi$-{\em compass}
if $\phi \geq |\phi_i(t)|$ for every $i \in \{0,1\}$ and all $t \in \Natural$,
i.e, a compass such that the absolute value of the deviation angle 
is bounded by~$\phi$.
A compass is said to be {\em static} if $\phi_i(t)$ is constant in $t$.
A compass is said to be {\em dynamic} if it can change its bearings at any
time~$t \in \Natural$, prior to the look phase.
A $\phi$-static compass is a $\phi$-compass that is static
and a $\phi$-dynamic compass is a $\phi$-compass that is dynamic.

\bigskip
\noindent
{\bf Gathering Problem:}~~

Let ${\cal L} = \{ (\mathbf{p},\mathbf{p}):~\mathbf{p} \in \Real^2 \}$
be the set of all configurations in which two robots are co-located.
An execution ${\cal E}: C(0), C(1), \ldots$ is called a {\em gathering execution},
if there are a configuration $C \in {\cal L}$ and a time instant~$f \in \Natural$ 
such that $C(t) = C$ holds for all $t \geq f$.
An algorithm is said to be a {\em gathering algorithm},
if for any configuration $C(0)$, 
every execution ${\cal E}: C(0), C(1), \ldots$ with initial configuration $C(0)$
is a gathering execution.

An algorithm is given as a function and is deterministic.
Nevertheless, execution $\cal E$ is not uniquely determined
from a given initial configuration $C(0)$.
Execution~$\cal E$ varies depending on many factors,
e.g., when each robot is activated,
how and when each scale ratio and compass change,
how far each robot moves, and so on.
We consider that all of these factors are controlled 
by an adversary playing against the gathering algorithm.
This paper investigate the problem of designing a gathering algorithm.

\bigskip
\noindent
{\bf Noetherian Termination}~~

The definition of a gathering algorithm is based on the Noetherian termination.
It does not request a gathering algorithm to eventually terminate.
All gathering algorithms that we present in this paper are of this type.
A stronger (and perhaps more conventional) definition additionally imposes 
the termination condition upon a gathering algorithm.
We will observe that the gathering algorithms for semi-synchronous
robots in Section~\ref{Ssemisynch} are transformable into 
gathering algorithms satisfying the termination condition.\footnote{
This transformation is not applicable to the gathering algorithms 
for asynchronous robots in Section~\ref{Sasynch}.}


\section{Semi-Synchronous Robots with Compasses}
\label{Ssemisynch}

In this section,
we investigate the gathering problem for two oblivious semi-synchronous
anonymous mobile robots with static and then dynamic compasses.
By definition, if the problem is solvable for the robots 
with $\phi$-dynamic compasses, 
it is also solvable for the robots with $\phi$-static compasses.
We establish, for each of the static and the dynamic cases,
the tight bound on $\phi$ for the problem to become solvable.

Consider, for an algorithm,
a finite execution ${\cal E}: C(0), C(1), \ldots, C(f)$ 
with an initial configuration $C(0)$,
and an execution ${\cal E'}: C'(0), C'(1), \ldots$ 
with an initial configuration $C'(0)$.
If $C'(0) = C(f)$,
then the concatenation ${\cal EE'}$ of ${\cal E}$ and ${\cal E'}$,
i.e.,  $C(0), C(1), \ldots,$ $C(f) (= C'(0)), C'(1), \ldots$,
is an execution with initial configuration $C(0)$,
since the robots are semi-synchronous.\footnote{
Asynchronous robots do not have this property,
since a robot in $\cal E$ may still be engaged in its move phase at time~$f$.}
In this section, we implicitly rely on this property.

\subsection{Semi-Synchronous Robots with Static Compasses}
\label{SSsemisynchstatic}

We now investigate the static case, i.e., 
the gathering problem for two oblivious robots with static compasses
under the semi-synchronous model.
Since the compasses are static, let $\phi_i(t) = \phi_i$ for $i \in \{0,1\}$.
The following theorem is a restatement of Theorem~3.1 of~\cite{SY99}.

\begin{thm}\cite{SY99}
\label{TSSnecessity}
There is no gathering algorithm for two oblivious anonymous 
robots with $\pi/2$-static compasses, 
under the semi-synchronous model.
\end{thm}

We present an algorithm $A_{SS}^{\phi}$\footnote{$SS$ of $A_{SS}^{\phi}$
stands for Semi-synchronous robots with Static compasses.}, and
show that it is a correct gathering algorithm for two oblivious
semi-synchronous robots with $\phi$-static compasses, provided
that $0 \leq \phi < \pi/2$.
Recall that an algorithm~$\psi$ is a total function on~$\Real^2$.
For any $\mathbf{p} = (u,v) \in \Real^2 \setminus \{ \mathbf{0} \}$,
let $\argum(\mathbf{p}) = \omega$ be the argument (or phase) of $\mathbf{p}$, 
i.e., $0 \leq \omega < 2\pi$ and 
$(u,v) = |\mathbf{p}| ( \cos \omega , \sin \omega)$.
Angles are calculated modulo~$2\pi$ in the sequel.

\begin{description}
\item[Algorithm $A_{SS}^{\phi}(\mathbf{p})$]~~

\begin{description}
\item[G(athered):]
If $\mathbf{p} = \mathbf{0}$ then $A_{SS}^{\phi}(\mathbf{p}) = \mathbf{0}$.
\item[A(pproach):]
If $0 < \argum(\mathbf{p}) \leq \pi$ 
then $A_{SS}^{\phi}(\mathbf{p}) = \mathbf{p}$.
\item[R(otate):]
If $\pi < \argum(\mathbf{p}) \leq 3\pi/2 + \phi$  
then $A_{SS}^{\phi}(\mathbf{p}) = (-|\mathbf{p}|,0)$.
\item[W(ait):]
If $3\pi/2 + \phi < \argum(\mathbf{p}) \leq 2\pi$  
then $A_{SS}^{\phi}(\mathbf{p}) = \mathbf{0}$.
\end{description}
\end{description}

\begin{figure}
\begin{center}
\includegraphics[keepaspectratio,width=70mm]{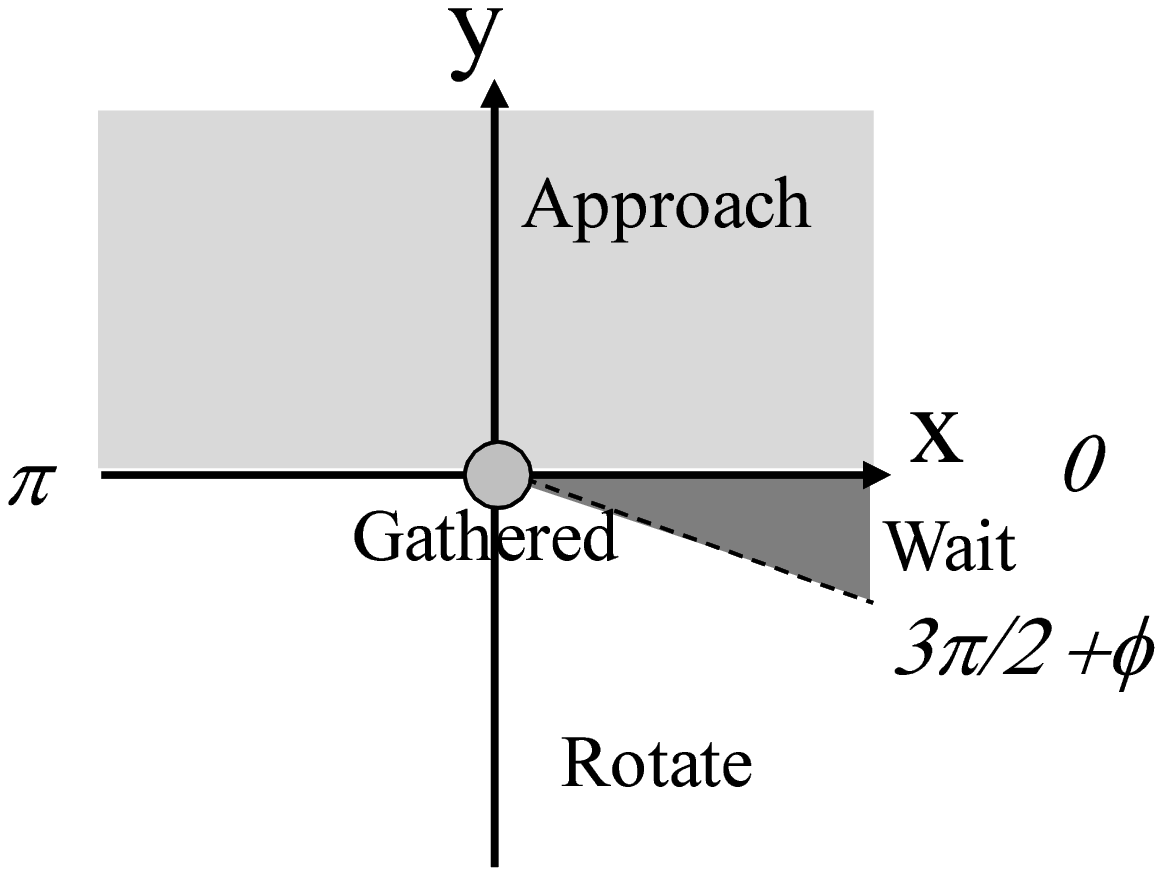}
\caption{An illsutration of Algorithm $A_{SS}^{\phi}$.}
\label{fig:SScoloredalg}
\end{center}
\end{figure}

\begin{figure}
\begin{center}
\includegraphics[keepaspectratio,width=70mm]{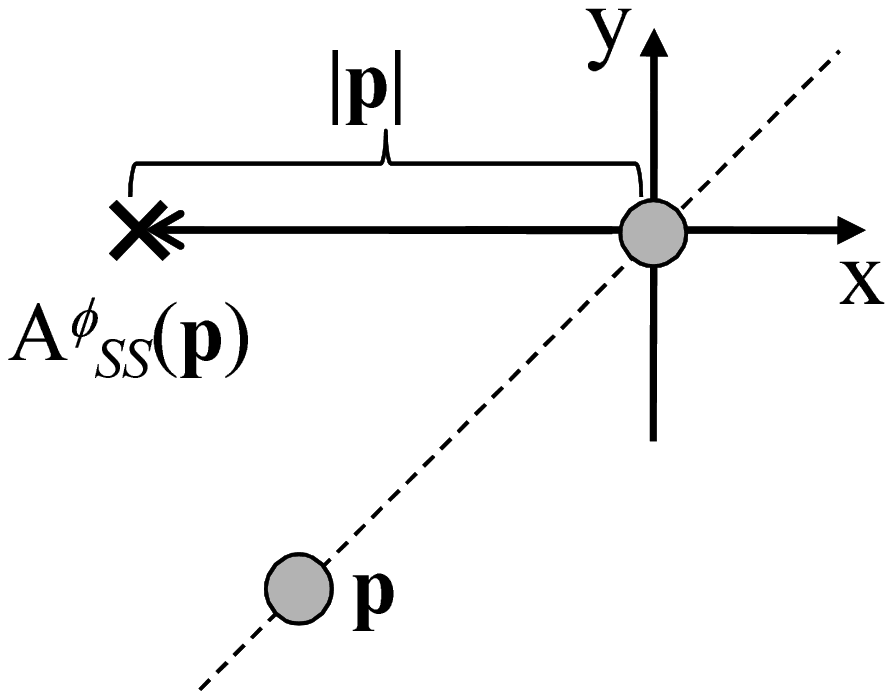}
\caption{The move of a robot  state Rotate who looks the other
robot at $\mathbf{p}$ in its local coordinate system.}
\label{fig:SSrotation}
\end{center}
\end{figure}

Fig.~\ref{fig:SScoloredalg} illustrates Algorithm 
$A_{SS}^{\phi}(\mathbf{p})$.
It dvides the plane 
into four regions Gathered, Approach, Rotate, and
Wait, and asks a robot to take the move corresponding to the region to 
which the current position $\mathbf{p}$ of the other robot
belongs. Each region is specified by the angle of its boundary
(except Gathered, whose region is a singleton
$\{ \mathbf{0}\}$). For example, in the case of Approach, the 
corresponding region is specified by two boundary angles $0$ and $\pi$.
A robot moves toward the other robot in state~$A$ (Approach), it
moves westward (i.e., negative direction) 
on its local $x$-axis in state~$R$ (Rotate),
and stops in state $W$ (Wait). 
The robots in state $G$ (Gathered) state are co-located,
and no further actions are necessary.
Although the actions at $A$, $W$ and $G$ are intuitive,
the action at~$R$ is less straightforward.
Figure \ref{fig:SSrotation} illustrates the move of
a robot in state Rotate who looks the other robot at 
$\mathbf{p}$ in its local coordinate system. We shall 
observe in more details how it rotates,
which is the core of $A_{SS}^{\phi}$.
In general, the state of a robot depends 
both on its current local coordinate system and the current position of the other robot.

Let $C(0)$ be a configuration and ${\cal E}: C(0), C(1), \ldots$ 
be an execution (of $A_{SS}^{\phi}$ on $\cal R$)
with initial configuration~$C(0)$,
where $C(t) = (\mathbf{r}_0(t), \mathbf{r}_1(t))$ is 
the configuration at time instant $t$, 
i.e., $\mathbf{r}_i(t)$ ($i \in \{0,1\}$)
is the location of $r_i$ in $Z$ at time~$t$.
The {\em state pair} $S(C(t))$ of configuration $C(t)$ 
at time $t$ is a pair $(s_0,s_1)$,
where $s_i$ ($i \in \{0,1\}$) is the state of robot $r_i$ at time $t$.
As mentioned,  $s_i$ may depend both on $C(t)$ and $Z_{(i,t)}$.

First of all, we confirm that state~$G$ corresponds to a ``goal'' configuration.
Suppose that the state of a robot, say $r_0$, is $G$ at time $t$.
Since $Z_{(0,t)}(\mathbf{r}_1(t)) = \mathbf{0}$,
$\mathbf{r}_0(t) = \mathbf{r}_1(t)$ 
and the state of $r_1$ at $t$ is also $G$.
Since they do not move in the time interval $[t,t+1]$
regardless of whether they are activated or not at $t$,
we obtain that $C(t') = C(t) \in {\cal L}$ for all $t' \geq t$.
We can thus conclude that $A_{SS}^{\phi}$ is correct 
if there is a time instant $t \in \Natural$ such that $S(C(t)) = (G,G)$.

Since the state of a robot is $G$
if and only if the state of the other robot is $G$, then
$S(C(t)) \not\in \{(G,s), (s,G)| s \in \{A, R, W\} \}$.
By the definition of $A_{SS}^{\phi}$,
$S(C(t)) = (W,W)$ obviously never occurs;
the execution never reaches a deadlock configuration $C(t)$
such that $S(C) = (W,W)$, in which neither robots move.

We next examine the case in which $\cal E$ reaches
a configuration $C(t)$ such that $S(C(t)) \in \{(W,A), (A,W)\}$.
The robot in state~$A$, say $r_0$, moves toward $r_1$,
while $r_1$, in state~$W$, stays motionless in time interval $[t,t+1]$.
By definition, if the distance between $r_0$ and $r_1$ is $\delta$ or less, 
then $r_0$ has reached the position of $r_1$ by $t+1$ 
and $S(C(t+1)) = (G,G)$ holds.
If $r_0$ has not reached the position of $r_1$ at $t+1$,
$S(C(t+1)) = S(C(t)) = (A,W)$ 
and the distance between $r_0$ and $r_1$ is now shorter,
since the position of $r_0$ at $t+1$ lies
on the line segment $\overline{\mathbf{r}_0(t)\mathbf{r}_1(t)}$.
We thus conclude that $A_{SS}^{\phi}$ is correct 
if there is a time instant $t \in \Natural$
such that $S(C(t)) \in \{(W,A), (A,W)\}$.

We have already shown that $A_{SS}^{\phi}$ is correct
if $S(C(0)) \in \{ (G,G), (A,W), (W,A)\}$.
If $S(C(0))$ is in none of $(G,G)$, $(A,W)$, $(W,A)$, then
the robots ``rotate'' the line segment 
$\overline{\mathbf{r}_0(t)\mathbf{r}_1(t)}$ counterclockwise 
until a state pair of either $(A,W)$ or $(W,A)$ occurs.
This task is done by robots in $R$ (i.e., Rotate) state.
The rest of this subsection is essentially devoted to showing this.
We now summarize some of the basic properties observed above.

\begin{property}
\label{PSSbasicproperties}~~

\begin{enumerate}
\item
State pair $(G,G)$ corresponds to a goal configuration.

\item
For any configuration $C(t)$,
$S(C(t)) \in \{(G,G), (A,A), (R,R), (A,R), (A,W), (R,A), (R,W), $
$(W,A), (W,R)\}$.
\item
If a configuration $C(t)$ such that $S(C) \in \{(A,W),(W,A)\}$ is reached,
then a goal configuration $(G,G)$ will be reached eventually.
\end{enumerate}
\end{property}

\begin{lma}
\label{LSSparallel}
Suppose that $\phi_0 = \phi_1$.
Then $A_{SS}^{\phi}$ correctly solves the gathering problem
for two oblivious robots under the semi-synchronous model.
\end{lma}

\begin{figure}
\begin{center}
\includegraphics[keepaspectratio,width=85mm]{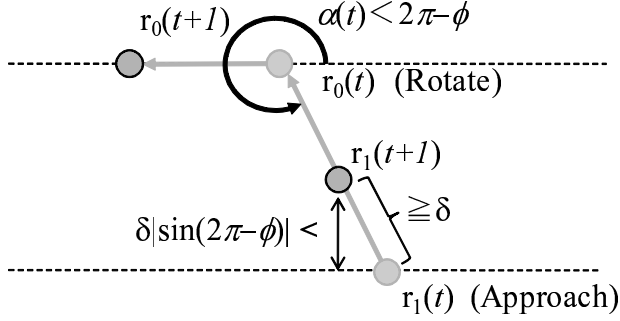}
\caption{An illustration used in the proof of Lemma~\ref{LSSparallel}.}
\label{fig:SSparallel}
\end{center}
\end{figure}

\begin{pf}
Recall that $C(t) = (\mathbf{r}_0(t), \mathbf{r}_1(t))$ is the
configuration at time $t$.
The coordinates of the position of robot $r_i$ ($i \in \{ 0,1\}$)
in $Z$ at $t$ are denoted by $\mathbf{r}_i(t) = (x_i(t), y_i(t))$.
By Property~\ref{PSSbasicproperties}, it suffices to show that ${\cal E}$ eventually reaches 
a configuration~$C(f)$ such that $S(C(f)) \in \{ (G,G), (A,W), (W,A)\}$ 
for some $f \in \Natural$.

The state pair $S(C(0))$ of initial configuration $C(0)$
must contain $A$ as the state of a robot because $\phi_0 = \phi_1$,
and there is nothing to show if $S(C(0)) \in \{ (G,G), (A,W), (W,A)\}$.
Hence we need to prove the lemma only for configurations $C(0)$ such that
$S(C(0)) \in  \{ (A,R), (R,A)\}$.
Without loss of generality, we assume the following:
\begin{enumerate}
\item
$\phi_0 = \phi_1 = 0$,
\item
$S(C(0)) = (R,A)$,
\item
$\mathbf{r}_0(0) = (x_0(0), y_0(0)) = \mathbf{0}$,
i.e., the position of $r_0$ is at the origin in $Z$, and 
\item
$y_1(0) < 0$.\footnote{
Observe that the state pair of $C(0)$ is either 
$(G,G), (A,W)$ or $(W,A)$ if $y_1(0) = 0$.}
\end{enumerate}

We assume that ${\cal E}$ never reaches a configuration~$C(f)$ 
such that $S(C(f)) \in \{ (G,G), (A,W), (W,A)\}$ and derive a contradiction.
Because $x_0(1) \leq 0, y_0(1) = 0$ and $y_1(1) \leq 0$,
$S(C(1)) \in \{ G,G), (W,A), (R,A)\}$,
which implies that $S(C(1)) = (R,A)$,
since $S(C(1)) \not\in \{ (G,G), (W,A)\}$.
Hence $S(C(t)) = (R,A)$ holds for all $t \in \Natural$.

Note that $x_0(t) \leq 0, y_0(t) = 0$, and $y_1(t) \leq 0$ hold.
Let $\alpha(t) = \argum(Z_{(0,t)}(\mathbf{r}_1(t)))$.
By definition, $\pi < \alpha(t) \leq 2\pi - \phi$.
Let $T = \{t_1, t_2, \ldots\}$ be the time instants 
at which $r_0$ is activated.
$T$ is an infinite set, since the execution is fair.
Obviously, by definition $\alpha(t_i) < \alpha(t_i + 1)$ 
for any $i \in \Natural$,
which implies that $\alpha(t)$ converges to 
an angle $\alpha \leq 2\pi - \phi$.
By the definition of state $R$, $3\pi/2 \leq \alpha$.

We now derive a contradiction.
Let $U = \{u_1, u_2, \ldots\}$ be the time instants 
at which $r_1$ is activated.
$U$ is also an infinite set, since the execution is fair.
Robot $r_1$ is in state $A$ at any time instant $u_i \in U$.
If $r_1$ reaches the next position at $u_i + 1$,
$y_1(u_i+1) = 0$ and the state pair of $C(u_i + 1)$
is either $(G,G)$ or $(W,A)$, a contradiction.
Thus, it moves by at least $\delta$ in time interval $[u_i,u_i + 1]$.
Since $3\pi/2 \leq \alpha(u_i) < 2\pi -  \phi$,
$y_1(u_i + 1) - y_1(u_i) > \delta |\sin (2\pi - \phi)|$ 
(see Fig.~\ref{fig:SSparallel} for an illustration).
It is a contradiction, since $y_1(t) \leq 0$ for any $t \in \Natural$.
\hfill $\Box$
\end{pf}

\begin{lma}
\label{LSSgeneral}
Suppose that $\phi_0 \not= \phi_1$.
Then, $A_{SS}^{\phi}$ correctly solves the gathering problem 
for two oblivious robots under the semi-synchronous model.
\end{lma}

\begin{pf}
It suffices to show,  by Property~\ref{PSSbasicproperties},  that ${\cal E}$ eventually reaches 
a configuration~$C(f)$ such that $S(C(f)) \in \{ (G,G), (A,W), (W,A)\}$ 
for some $f \in \Natural$.
Since $\phi_0 \not= \phi_1$,  
we assume $\phi_0 < \phi_1$ without loss of generality.
Since $\phi < \pi/2$, 
the angle formed by their $x$-axes is less than $\pi$.
Consider $C(t)$ for any $t \in \Natural$.
Let $\mathbf{o}(t)$ be the intersection of the $x$-axes 
of $Z_{(0,t)}$ and $Z_{(1,t)}$.
We may assume that a robot is not on the $x$-axis of the other at $t$,
since $S(C(t)) \not\in \{ (G,G), (A,W), (W,A)\}$.
Let $Z_{(i,t)}(\mathbf{p}) = (x_{(i,t)}(\mathbf{p}),y_{(i,t)}(\mathbf{p}))$
for $i \in \{0,1\}$ and $\mathbf{p} \in \Real^2$.
By definition, for $i  \in \{0,1\}$,
$x_{(i,t)}(\mathbf{o}(t))$ and $x_{(i,t)}(\mathbf{r}_i(t))$
are the $x$-coordinates, in $Z_{(i,t)}$ at time $t$,
of the intersection $\mathbf{o}(t)$
and the position of $r_i$, respectively.
According to their relative positions on the $x$-axis of $Z_{(i,t)}$,
we partition the configurations that may occur in $\cal E$
into four classes:
\begin{description}
\item[P(ositive)P(ositive):]
$x_{(0,t)}(\mathbf{o}(t)) < x_{(0,t)}(\mathbf{r}_0(t))$ 
and $x_{(1,t)}(\mathbf{o}(t)) < x_{(1,t)}(\mathbf{r}_1(t))$ 
\item[P(ositive)N(egative):]
$x_{(0,t)}(\mathbf{o}(t)) < x_{(0,t)}(\mathbf{r}_0(t))$ 
and $x_{(1,t)}(\mathbf{o}(t)) > x_{(1,t)}(\mathbf{r}_1(t))$ 
\item[N(egative)P(ositive):]
$x_{(0,t)}(\mathbf{o}(t)) > x_{(0,t)}(\mathbf{r}_0(t))$ 
and $x_{(1,t)}(\mathbf{o}(t)) < x_{(1,t)}(\mathbf{r}_1(t))$ 
\item[N(egative)N(egative):]
$x_{(0,t)}(\mathbf{o}(t)) > x_{(0,t)}(\mathbf{r}_0(t))$ 
and $x_{(1,t)}(\mathbf{o}(t)) > x_{(1,t)}(\mathbf{r}_1(t))$ 
\end{description}
Fig.~\ref{fig:four_cases} illustrates these four cases.
In the following, 
we show that ${\cal E}$ eventually reaches 
a configuration~$C(f)$ such that $S(C(f)) \in \{ (G,G), (A,W), (W,A)\}$, 
for each of the four cases to which $C(0)$ may belong: PP, PN, NP and NN.

\begin{figure}
\begin{center}
\includegraphics[keepaspectratio,width=150mm]{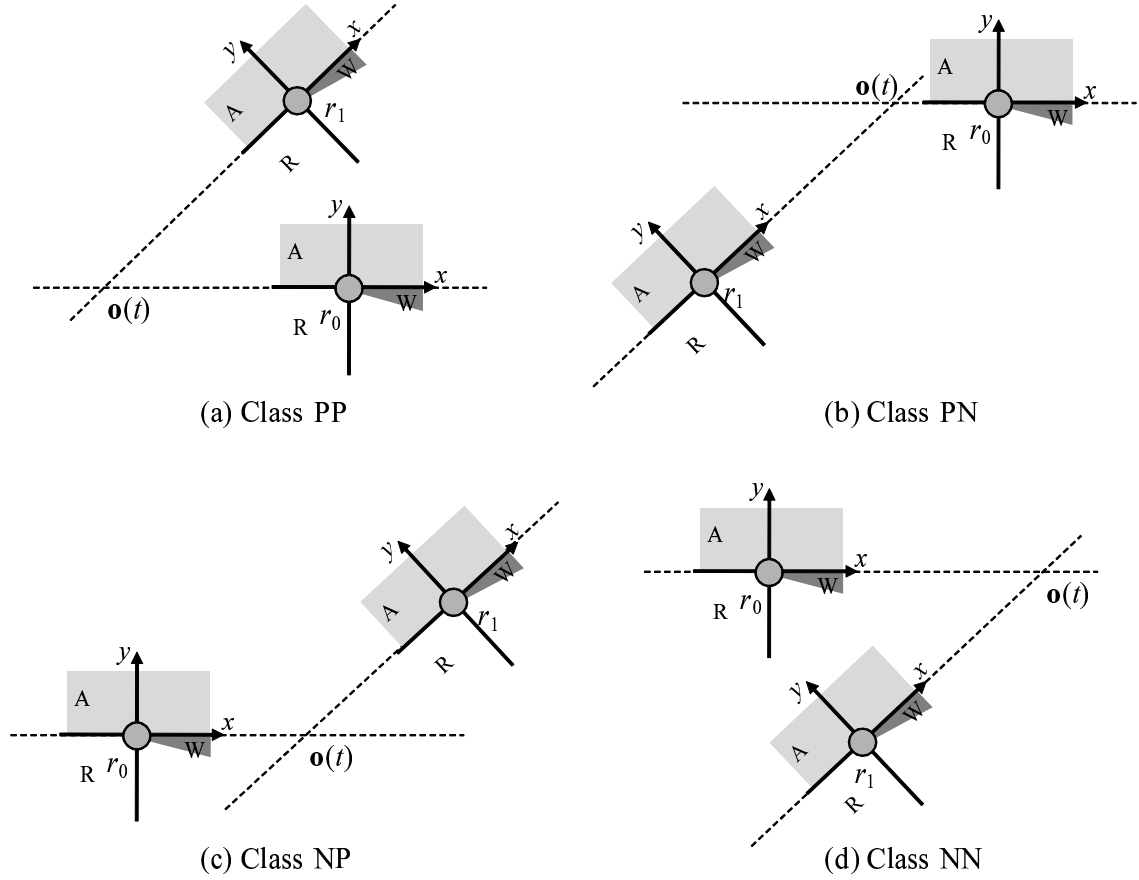}
\caption{Illustrations of the four cases PP, PN, NP and NN
used in the proof of Lemma~\ref{LSSgeneral}.}
\label{fig:four_cases}
\end{center}
\end{figure}

\noindent
{\bf Case NN}:
Suppose that $C(0)$ is in class NN,
which implies that $S(C(0)) = (R,A)$.
In this case, by using an argument similar to the one used in the proof
of Lemma~\ref{LSSparallel}, 
we can show that $\cal E$ eventually reaches 
a configuration $C(f)$ such that $S(C(f)) \in \{(G,G), (W,A)\}$.

\noindent
{\bf Case PN}:
Suppose that $C(0)$ is in class PN,
which implies that $S(C(0)) \in \{(R,R), (R,W)\}$.
Since $r_0$ goes west and $r_1$ goes west or stays motionless 
and thus $r_0$ decreases its $x$-coordinate and $r_1$ does 
not increase its $x$-coordinate
(without changing their $y$-coordinates)
in their local coordinate systems,
$\cal E$ eventually reaches a configuration $C(f)$ in class NN.
Thus, this case is reduced to Case NN.

\noindent
{\bf Case NP}:
Suppose that $C(0)$ is in class NP,
which implies that $S(C(0)) = (A,A)$.
Since the robots move toward each other's position,
$\cal E$ eventually reaches a configuration $C(f)$ in class PN,
unless it reaches a configuration $C(f)$ such that $S(C(f)) = (G,G)$ directly.
Thus, this case is reduced to Case PN.

\noindent
{\bf Case PP}:
Suppose finally that  $C(0)$ is in class PP,
which implies that $S(C(0)) = (A,R)$.
Since $r_0$ never reaches the $x$-axis of $Z_{(1,t)}$,
and $r_1$ decreases its $x$-coordinate in $Z_{(1,t)}$, it follows that
$\cal E$ eventually reaches a configuration $C(f)$ in class PN.
Thus, this case is also reduced to Case PN.
\hfill $\Box$
\end{pf}

By Lemmas~\ref{LSSparallel} and \ref{LSSgeneral},
we derive the following theorem.

\begin{thm}
\label{TSSsufficency}
For any $0 \leq \phi < \pi/2$,
Algorithm $A_{SS}^{\phi}$ for two oblivious robots 
that uses $\phi$-static compasses solves the
gathering problem, under the semi-synchronous model.
\end{thm}

\subsection{Semi-Synchronous Robots with Dynamic Compasses}
\label{SSsemisynchdinamic}

We investigate the gathering problem for two semi-synchronous
robots with $\phi$-dynamic compasses for some $\phi$.
Unlike the previous subsection,
$\phi_i(t)$ can now vary in time, as long as $|\phi_i(t)| \leq \phi$ always holds.

The proof of Theorem~3.1 of \cite{SY99} shows that an algorithm 
solves the gathering problem under the semi-synchronous model,
only if there is a configuration such that one robot,
say $r_0$, moves to the position of $r_1$ while $r_1$ stays motionless.
We can restate this condition, using the notation introduced
in Subsection~\ref{SSsemisynchstatic}, as follows:
A gathering algorithm is correct, 
only if there is a configuration $C$ such that 
the corresponding state pair $S(C)$ is either $(A,W)$ or $(W,A)$.
We say that a configuration $C$ is {\em stable} 
if $S(C)$ is determined uniquely,  
regardless of the current local coordinate systems $Z_{(i,t)}$.
Following the proof of Theorem~3.1 in \cite{SY99}
for semi-synchronous robots, and
additionally taking into account that they have dynamic compasses,
we have the following property.

\begin{property}
\label{PSDstable}
An algorithm solves the gathering problem for two oblivious
semi-synchronous robots with dynamic compasses,
only if there is a stable configuration $C$ such that 
$S(C)$ is either $(A,W)$ or $(W,A)$.
\end{property}

\begin{thm}
\label{TSDnecessity}
There is a gathering algorithm for two oblivious semi-synchronous 
robots with $\phi$-dynamic compasses, only if $\phi < \pi/4$.
\end{thm}

\begin{pf}
It suffices to show that there is 
no gathering algorithm for $\phi = \pi/4$.
We assume that such an algorithm exists, called~$ALG$, in order 
to derive a contradiction.
Then, by Property~\ref{PSDstable}, there is a stable configuration $C$ such that
the corresponding state pair $S(C)$ is either $(A,W)$ or $(W,A)$.
We assume, without loss of generality, that $C = ((0,0),(1,0))$ and $S(C) = (W,A)$.
Since $C$ is stable and $\phi_{0}(t)$ can be $\pm \pi/4$, we have that
$S(C') = S(C'') = (W,A)$,
where $C' = ((0,0),(\sqrt{2}/2,\sqrt{2}/2))$,
and $C'' = ((0,0),(\sqrt{2}/2,-\sqrt{2}/2))$.

Consider an execution starting with initial configuration 
$C(0) = ((0,0),(0,1))$,
assume that $\phi_{0}(0) = \pi/4$ and $\phi_{1}(0) = -\pi/4$.
Then $S(C(0)) = (W,W)$, a contradiction.
\hfill $\Box$
\end{pf}

We next present an algorithm $A_{SD}^{\phi}$ and show that 
it is a correct gathering algorithm for two oblivious semi-synchronous
robots with $\phi$-dynamic compasses,
provided that $0 \leq \phi < \pi/4$.\footnote{$SD$ of $A_{SD}^{\phi}$
stands for Semi-synchronous robots with Dynamic compasses.}
For any $\mathbf{p} \in \Real^2$ and angle $\omega$,
Let $\rho_{\omega}(\mathbf{p}) = \mathbf{q}$, where
\[
\mathbf{q}^T = \left(
\begin{array}{rr}
\cos \omega & - \sin \omega\\
\sin \omega & \cos \omega \\ 
\end{array}
\right) \mathbf{p}^T,
\]
that is, $\rho_{\omega}(\mathbf{p})$ is the point obtained by
rotating $\mathbf{p}$ by angle $\omega$ 
with respect to the rotation center $\mathbf{0}$.

\begin{description}
 \item[Algorithm $A_{SD}^{\phi}(\mathbf{p})$]~~

\begin{description}
\item[G(athered):]
If $\mathbf{p} = \mathbf{0}$ 
then $A_{SD}^{\phi}(\mathbf{p}) = \mathbf{0}$.
\item[A(pproach):]
If $\pi/2 + \phi < \argum(\mathbf{p}) \leq 3\pi/2 - \phi$
then $A_{SD}^{\phi}(\mathbf{p}) = \mathbf{p}$.
\item[W(ait):]
If $- \pi/2 + \phi < \argum(\mathbf{p}) \leq \pi/2 - \phi$ 
then $A_{SD}^{\phi}(\mathbf{p}) = \mathbf{0}$.
\item[R(otate):]
If $\pi/2 - \phi < \argum(\mathbf{p}) \leq \pi/2 + \phi$
or $3\pi/2 - \phi < \argum(\mathbf{p}) \leq 3\pi/2 + \phi$
$(= - \pi/2 + \phi)$, 
then $A_{SD}^{\phi}(\mathbf{p}) = \rho_{\frac{\pi}{2} + \phi}(\mathbf{p})$. 
\end{description}
\end{description}

\begin{figure}
\begin{center}
\includegraphics[keepaspectratio,width=70mm]{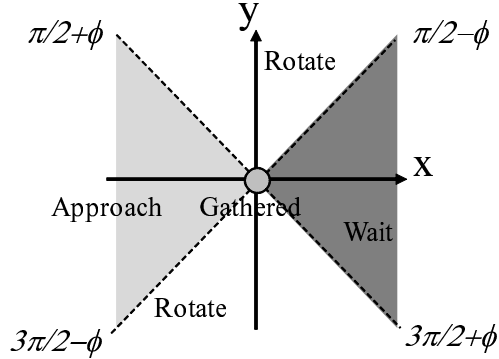}
\caption{An illustration of Algorithm $A_{SD}^{\phi}$.}
\label{fig:SDcoloredalg}
\end{center}
\end{figure}

\begin{figure}
\begin{center}
\includegraphics[keepaspectratio,width=40mm]{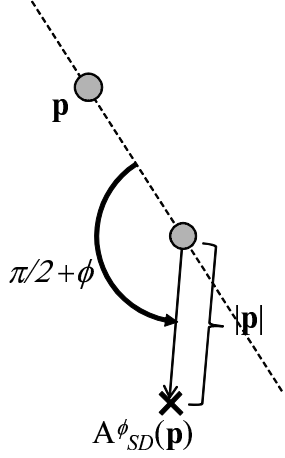}
\caption{The move of a robot in state Rotate who looks the other robot
at $\mathbf{p}$ in its local coordinate system.}
\label{fig:SDrotation}
\end{center}
\end{figure}

An illustration of this algorithm is shown in 
Fig.~\ref{fig:SDcoloredalg}.
Like $A_{SS}^{\phi}$, 
a robot moves toward the other robot in state $A$ (i.e., Approach)
and stays there motionless in state $W$ (i.e, Wait).
Although the action taken in state $R$ (i.e., Rotate) seems 
slightly more complex than for $A_{SS}^{\phi}$,
the idea behind the definition is similar to $A_{SS}^{\phi}$. 
This additional complexity, illustrated in Fig.~\ref{fig:SDrotation}, comes 
from the need to handle the dynamic compasses.
Roughly, a robot at $R$ rotates the line segment 
connecting the current robots' positions clockwise,
until its deviation from the $x$-axis of $Z$ becomes 
smaller than $\pi/2 - \phi (> 0)$.
Since such a configuration $C$ is stable and $S(C)$ is
either $(A,W)$ or $(W,A)$,
the gathering is eventually achieved.

Just like with $A_{SS}^{\phi}$, the
state pair $(G,G)$ corresponds to a goal (i.e., gathered) configuration.
Unlike with $A_{SS}^{\phi}$,
state pair $(A,A)$ never occurs, in addition to the state pairs
in $\{ G,s), (s,G), (W,W)| s \in \{A, R, W\} \}$.
Note that Item~3) of Property~\ref{PSSbasicproperties}
does not hold for  $A_{SD}^{\phi}$, i.e.,
not all configurations $C$, with $S(C) \in \{(A,W),(W,A)\}$, are stable.

We explain the intention of the definition
$A_{SD}^{\phi}(\mathbf{p}) = \rho_{\pi/2 + \phi}(\mathbf{p})$ of $R$.
Suppose that an execution reaches  at time $t$
a configuration $C = (\mathbf{r}_0, \mathbf{r}_1)$,
in which a robot, say $r_0$, is in state $R$.
For the simplicity of the explanation, 
assume that $\mathbf{r}_0 = \mathbf{0}$, 
and that $y_1 > 0$,  where $\mathbf{r}_1 = (x_1, y_1)$.
Since $r_0$ is at $R$,
$\pi/2 - \phi < \argum(Z_{(0,t)}(\mathbf{r}_1)) \leq \pi/2 + \phi$ 
in $Z_{(0,t)}$.
As $|\phi_0(t)| \leq \phi$,
$\pi/2 - 2\phi < \argum(\mathbf{r}_1) \leq \pi/2 + 2\phi$ in $Z$.
The direction $\theta$ of the next position hence 
satisfies $\pi < \theta < \argum(\mathbf{r}_1) + \pi (< 2\pi)$.
Since $\theta < \argum(\mathbf{r}_1) + \pi$, 
a robot at $R$, once activated, rotates the
line segment $\overline{\mathbf{r}_0\mathbf{r}_1}$ clockwise.
Since $\pi < \theta$,
the rotation of $\overline{\mathbf{r}_0\mathbf{r}_1}$
never exceeds the $x$-axis of $Z$.

\begin{thm}
\label{TSDsufficiency}
For any $0 \leq \phi < \pi/4$,
Algorithm $A_{SD}^{\phi}$ for two oblivious 
robots that use $\phi$-dynamic compasses solves the
gathering problem, under the semi-synchronous model.
\end{thm}

\begin{figure}
\begin{center}
\includegraphics[keepaspectratio,width=150mm]{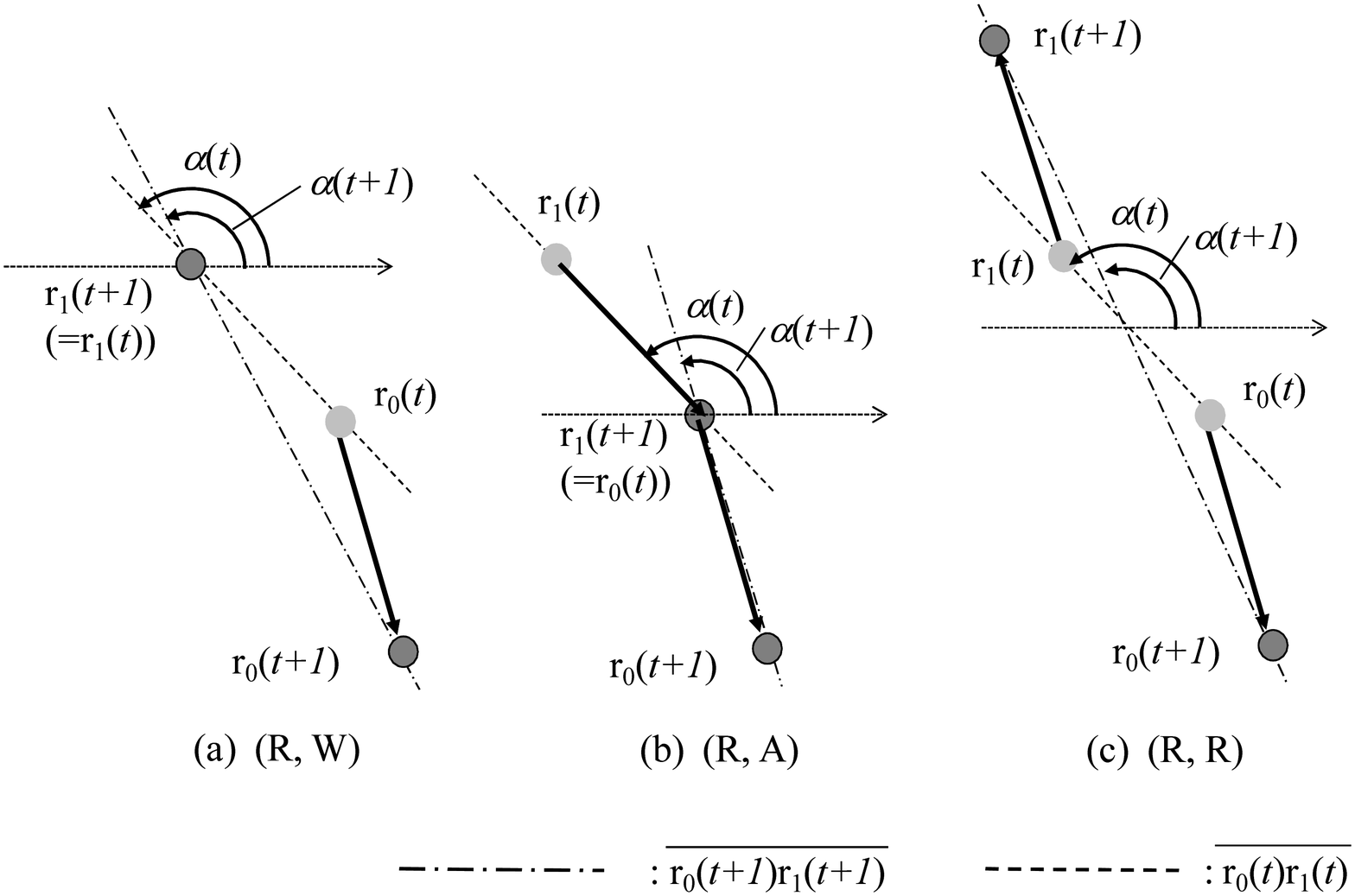}
\caption{Illustrations to explain why $\alpha(t)$ monotonically decreases,
which are used in the proof of Theorem~\ref{TSDsufficiency}.}
\label{fig:SDalphadec}
\end{center}
\end{figure}

\begin{figure}
\begin{center}
\includegraphics[keepaspectratio,width=120mm]{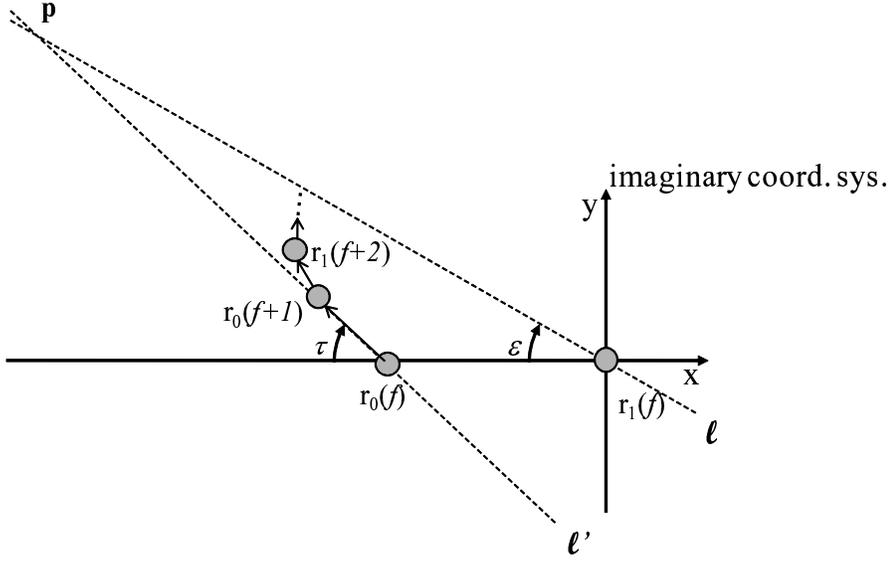}
\caption{An illustration of a contradictive situation in the proof of
Theorem~\ref{TSDsufficiency}.}
\label{fig:SDconvergence}
\end{center}
\end{figure}

\begin{pf}
It suffices to show that any execution ${\cal E} = C(0), C(1), \ldots$
eventually reaches a configuration $C(f)$ such that $S(C(f)) = (G,G)$.
We assume that there is a configuration $C(0)$
such that there is an execution ${\cal E} = C(0), C(1), \ldots$ 
in which $S(C(t)) \not= (G,G)$ holds for any $t$.
We then derive a contradiction.
Let $C(t) = (\mathbf{r}_0(t), \mathbf{r}_1(t))$,
where $\mathbf{r}_i(t) = (x_i(t),y_i(t))$ for $i \in \{0,1\}$,
and $\alpha(t) = \argum(\mathbf{r}_1(t) - \mathbf{r}_0(t))$.
If $y_1(0) = 0$ holds, it is a contradiction
since $C(0)$ is a stable configuration
such that $S(C(0))$ is either $(W,A)$ or $(A,W)$.
Without loss of generality, 
we thus assume that $\mathbf{r}_0(0) = \mathbf{0}$ and $y_1(0) > 0$.
Hence $0 <  \alpha(0) <  \pi$.

By the respective definitions of states $A$ and $R$,
and by the above observation about $R$, we obtain that
$0 < \alpha(t+1) \leq \alpha(t)$ for any $t \in \Natural$.
If a robot at $R$ is activated only a finite number of times,
then there is an infinite subexecution 
$C(f), C(f+1), \ldots$ such that $S(C(t))$ is
either $(A,W)$ or $(W,A)$ for any $t \geq f$ for some $f \in \Natural$---a contradiction.
A robot at $R$ is thus activated infinitely many times.
Since a robot at $R$ rotates 
segment $\overline{\mathbf{r}_0(t) \mathbf{r}_1(t)}$ clockwise 
whenever it is activated (Fig.~\ref{fig:SDalphadec}),
and $0 < \alpha(t)$ for all $t \in \Natural$, then
$\alpha(t)$ converges to an angle $\alpha > 0$.

We again derive a contradiction.
Since $\alpha(t)$ converges to $\alpha > 0$
for any small $\epsilon > 0$,
there is a time $f \in \Natural$ such that
$\alpha(t) - \alpha < \epsilon$ for all $t \geq f$.
Because an activation of $r_1$ does not increase $\alpha(t)$,
we may assume without loss of generality that only $r_0$ is activated after $f$.
Let $\tau = \pi/2 - \phi (= \pi - (\pi/2 + \phi))$.

For convenience,
imagine that $\mathbf{r}_1(f) = \mathbf{0}$ 
and $\mathbf{r}_0(f) = (-1,0)$.
Since $r_1$ is not activated after $f$,
$r_1$ stays at $\mathbf{0}$.
Let $\ell$ (resp. $\ell'$) be a half line ended
at $\mathbf{0}$ (resp. $(-1,0)$) with $\pi - \epsilon$
(resp. $\pi - \tau$) being the angle it makes with the $x$-axis,
and let $\mathbf{p}$ be the intersection of $\ell$ and $\ell'$ 
(see Fig.~\ref{fig:SDconvergence} for illustration).

Since $\tau \gg \epsilon$, $\mathbf{p}$ is in the second quadrant.
Let $X = \mathbf{r}_0(f) (= (-1,0)), \mathbf{r}_0(f+1), \ldots$
be the polygonal chain constructed from the positions of $r_0$ after $f$.
It is easy to observe that $X$ is entirely contained
within the triangle formed by vertices $\mathbf{0}, \mathbf{p}$ and $(-1,0)$. This is a contradiction with the fact that
$|\mathbf{r}_0(t+1) - \mathbf{r}_0(t)| \geq \delta$ and 
$\argum(\mathbf{r}_0(t+1) - \mathbf{r}_0(t)) \geq \pi - (\tau +  \epsilon)$,
for any $t \geq f$.
\hfill $\Box$
\end{pf}

\begin{remark}
Algorithms $A_{SS}^{\phi}$ and $A_{SD}^{\phi}$ run forever.
Let us modify them by replacing Gathered into the following
Terminate:
\begin{description}
\item[T(erminate):]
If $\mathbf{p} = \mathbf{0}$ then terminate.
\end{description}
The modified $A_{SS}^{\phi}$ and $A_{SD}^{\phi}$ then eventually terminate
at a gathered configuration because the original 
$A_{SS}^{\phi}$ and $A_{SD}^{\phi}$ have the following property:
For any execution ${\cal E}: C(0), C(1), \ldots$, 
there is a time instant $f \in \Natural$ such that $S(C(f)) = (G,G)$,
and for any $f \in \Natural$ such that $S(C(f)) = (G,G)$,
$S(C(t)) = (G,G)$ holds for any $t \geq f$.
\end{remark}

\section{Asynchronous Robots with Compasses}
\label{Sasynch}

We now address the case of asynchronous mobile robots.
As emphasized earlier, 
a main difference between asynchronous and semi-synchronous
robots is that in the former case,
a concatenation $\cal {EE'}$ of a finite execution $\cal E$
and an execution $\cal E'$ may not be a correct execution,
even if the last configuration $C(f)$ of $\cal E$ is 
the initial configuration of $\cal E'$. This is because, due to the asynchrony  of the three phases,
one of the two robots may be caught in the middle of its move phase in $C(f)$ in $\cal E$.
We say that a robot is {\em settled} at time $t$ 
if 1) it is not activated at $t$, or
2) it is activated at $t$ and
it will not change its position until it is next activated.
Obviously $\cal {EE'}$ is a correct execution,
if both robots are settled at time~$f$.

\subsection{Asynchronous Robots with Static Compasses}
\label{SSasynchstatic}

In Subsection~\ref{SSsemisynchstatic},
we presented Algorithm $A_{SS}^{\phi}$ 
and showed as Theorem~\ref{TSSsufficency} 
that it correctly solves the gathering problem 
for two oblivious robots using $\phi$-static compasses under
the semi-synchronous model, if $0 \leq \phi < \pi/2$.
We now show that Theorem~\ref{TSSsufficency} 
can be extended to asynchronous robots;
i.e., we show that $A_{SS}^{\phi}$ correctly solves the gathering
problem for two oblivious robots using $\phi$-static compasses 
under the asynchronous model, if $0 \leq \phi < \pi/2$.
We then conclude by Theorem~\ref{TSSnecessity} 
that there is a gathering algorithm
for two oblivious asynchronous robots using static compasses 
with maximum deviation $\phi$, if and only if $0 \leq \phi < \pi/2$.

We keep with the notation defined in Subsection~\ref{SSsemisynchstatic},
and follow the scenario that we adopted in the proof of Theorem~\ref{TSSsufficency}.
We show the correctness of $A_{SS}^{\phi}$ under the asynchronous model,
taking into account features that characterize an asynchronous execution.
Like semi-synchronous robots,
$S(C) \not\in \{(G,s), (s,G), (W,W)| s \in \{A, R, W\} \}$
for any configuration $C$.
Hence the execution never reaches a deadlock configuration $C$, 
in which neither robot can move, before reaching $(G,G)$.
However, unlike semi-synchronous robots,
state pair $(G,G)$ no longer characterizes a goal configuration,
since an asynchronous robot may still be unsettled.
In other words, it may be moving away without noticing 
that the gathering had just been completed.\footnote{
We call such a configuration a \emph{pseudo-gathered} configuration
in the next subsection.}

In order to handle such configurations $C(t)$ at which a robot is not settled,
we also pay attention to times $a_i(t)$ and $b_i(t)$ with $i \in \{ 0,1 \}$,
where $a_i(t)$ (resp. $b_i(t)$) is the last time before (and including) $t$ 
(resp. the first time after (and including) $t$) at which robot $r_i$ is activated.
If $r_i$ is activated at $t$, then $a_i(t) = b_i(t) = t$.
If $r_i$ is not activated at $t$,
then $r_i$ is not activated in time interval $(a_i(t), b_i(t))$.
Like Subsection~\ref{SSsemisynchstatic},
let us start with the simple case of $\phi_0 = \phi_1$.

\begin{lma}
\label{LASparallel}
Suppose that $\phi_0 = \phi_1$.
Then $A_{SS}^{\phi}$ correctly solves the gathering problem
for two oblivious robots under the asynchronous model.
\end{lma}

\begin{pf}
Our proof is very similar to that of Lemma~\ref{LSSparallel}.
Let $C(0)$ be any initial configuration.
Consider any execution ${\cal E}: C(0), C(1), \ldots$,
where $\mathbf{r}_i(t) = (x_i(t), y_i(t))$ 
for any $i \in \{ 0,1 \} $ and $t \in \Natural$.
Except that both states of robots are $G$, $S(C(0))$ must contain 
$A$ as the state of a robot. Thus, 
$S(C(0)) \in \{ (G,G), (A,R), (A,W), (R,A), (W,A)\}$ holds.

If $S(C(0)) = (G,G)$, 
then by the definition of $A_{SS}^{\phi}$,
$S(C(t)) = (G,G)$ for any $t \in \Natural$,
i.e., the gathering completes.

If $S(C(0)) \in \{ (A,W), (W,A)\}$,
then only the robot with state $A$, say $r_0$, 
can move (toward $r_1$) at $C(0)$,
and thus $S(C(1)) = (A,W)$.
Hence, ${\cal E}$ eventually reaches a configuration $C(t)$ 
(possibly after taking a number of configurations $C$ such that $S(C) = (A,W)$)
such that $S(C(t)) = (G,G)$.
Let $t_0$ be the earliest time instant $t$ at which $S(C(t)) = (G,G)$ holds.

We observe that both robots are settled.
The state of $r_1$ is $W$ at $a_1(t_0)$ and hence $r_1$ is settled at $t_0$.
Robot $r_0$ (whose state at $a_0(t_0)$ is $A$) is also settled at $t_0$
since, by the definition of $A_{SS}^{\phi}$, the next position of $r_0$ at $a_0(t_0)$ is the position of $r_1$.
It follows that gathering completes,
like the case where $S(C(0)) = (G,G)$.

We continue with the cases where $S(C(0)) \in \{ ((A,R), (R,A) \}$.
Like the proof of Lemma~\ref{LSSparallel},
assume without loss of generality that:
\begin{enumerate}
\item
$\phi_0 = \phi_1 = 0$,
\item
$S(C(0)) = (R,A)$,
\item
$\mathbf{r}_0(0) = (x_0(0), y_0(0)) = \mathbf{0}$,
i.e., the position of $r_0$ is at the origin in $Z$, and 
\item
$y_1(0) < 0$.\footnote{
The state pair of $C(0)$ is any of 
$(G,G), (A,W)$ or $(W,A)$ if $y_1(0) \geq 0$.}
\end{enumerate}

By the same argument used in the proof of Lemma~\ref{LSSparallel},
$y_1(t) \leq 0$ holds for any $t \in \Natural$.
If $y_1(t) = 0$ for some $t \in \Natural$, then
let $t_0$ be the earliest time instant $t$ at which $y_1(t) = 0$ holds.
It follows that $S(C(t_0)) \in \{(W,A),(G,G)\}$, and $r_1$ is settled at $t_0$
(because $\mathbf{r}_1(t_0)$ is the next position of $r_1$ at  $a_1(t_0)$).
If $r_0$ is settled at $t_0$, the gathering eventually completes 
since it reduces to the case where $S(C(0)) \in \{(W,A),(G,G)\}$.

If $r_0$ is not settled at $t_0$,
then $r_0$ is moving (or will move) in 
the negative direction on its $x$-axis,
since $S(C(a_0(t_0))) = (R,A)$.
Thus $r_0$ is settled at $b_0(t_0)$ and $S(C(b_0(t_0))) = (W,A)$.
If $r_1$ is settled at $b_0(t_0)$, 
then the gathering eventually completes as discussed above.
If $r_1$ is not settled at $b_0(t_0)$, 
then $S(C(t_1)) \in \{(W,A),(G,G)\}$ and
both $r_0$ and $r_1$ are settled at $t_1$,
where $t_1 = b_1(b_0(t_0))$.
Therefore, the gathering eventually completes.

To derive a contradiction, we next assume that $y_1(t) = 0$ does not hold 
for any $t \in \Natural$.
By the same argument used in the proof of Lemma~\ref{LSSparallel},
${\cal E}$ eventually reaches a configuration $C(t)$ such that $S(C(t)) = (W,A)$.
Let $t_0$ be the first time instant $t$ such that $S(C(t)) = (W,A)$ holds.
If both $r_0$ and $r_1$ are settled at $t_0$,
the gathering eventually completes, 
leading to a contradiction.
If $r_0$ is not settled and is moving (or will move)
in the negative direction on its $x$-axis at $t_0$,
the gathering eventually completes
by a similar argument as above.
A contradiction is thus derived.
\hfill $\Box$
\end{pf}

\begin{lma}
\label{LASgeneral}
Suppose that $\phi_0 \not= \phi_1$.
Then $A_{SS}^{\phi}$ correctly solves the gathering problem 
for two oblivious robots under the asynchronous model.
\end{lma}
 
\begin{pf}
Again, our proof is similar to that of Lemma~\ref{LSSgeneral}.
We continue to use the same concepts and notations,
but introduce them again for the convenience of the reader.

Consider any configuration $C(0)$ and any execution 
${\cal E} = C(0), C(1), \ldots$ starting at $C(0)$,
where $C(t) = (\mathbf{r}_0(t), \mathbf{r}_1(t))$ 
for any $t \in \Natural$.
We assume $\phi_0 < \phi_1$ without loss of generality.
Since $\phi < \pi/2$, 
we denote by $\mathbf{o}(t)$ the intersection of the $x$-axes 
of $Z_{(0,t)}$ and $Z_{(1,t)}$.
Let $Z_{(i,t)}(\mathbf{p}) = (x_{(i,t)}(\mathbf{p}),y_{(i,t)}(\mathbf{p}))$
for any $i \in \{ 0,1 \}$ and $\mathbf{p} \in \Real^2$.
By definition,
$x_{(i,t)}(\mathbf{o}(t))$ and $x_{(i,t)}(\mathbf{r}_i(t))$
are the $x$-coordinates, in $Z_{(i,t)}$ at time $t$,
of the intersection $\mathbf{o}(t)$ 
and the position of $r_i$, respectively.

As explained in the proof of Lemma~\ref{LSSgeneral},
under the semi-synchronous model,
we could assume without loss of generality that 
a robot is not on the $x$-axis of the other at $t$. Unfortunately, in the asynchronous model,
we can no longer assume this.
That is, a robot can possibly be located at $\mathbf{o}(t)$ at $t$.
Taking this into account,
we partition the configurations into four classes as follows.
(The partition is slightly different from the one 
defined in the proof of Lemma~\ref{LSSgeneral}.)

\begin{description}
\item[P(ositive)P(ositive):]
$x_{(0,t)}(\mathbf{o}(t)) < x_{(0,t)}(\mathbf{r}_0(t))$ 
and $x_{(1,t)}(\mathbf{o}(t)) < x_{(1,t)}(\mathbf{r}_1(t))$ 
\item[P(ositive)N(egative):]
$x_{(0,t)}(\mathbf{o}(t)) < x_{(0,t)}(\mathbf{r}_0(t))$ 
and $x_{(1,t)}(\mathbf{o}(t)) \geq x_{(1,t)}(\mathbf{r}_1(t))$ 
\item[N(egative)P(ositive):]
$x_{(0,t)}(\mathbf{o}(t)) \geq x_{(0,t)}(\mathbf{r}_0(t))$ 
and $x_{(1,t)}(\mathbf{o}(t)) < x_{(1,t)}(\mathbf{r}_1(t))$ 
\item[N(egative)N(egative):]
$x_{(0,t)}(\mathbf{o}(t)) \geq x_{(0,t)}(\mathbf{r}_0(t))$ 
and $x_{(1,t)}(\mathbf{o}(t)) \geq x_{(1,t)}(\mathbf{r}_1(t))$ 
\end{description}

\noindent
({\bf Case NN})
Suppose that $C(0)$ is in class NN,
which implies that $S(C(0)) \in \{(W,A), (R,A), (G, G)\}$.
We can show that gathering eventually completes in the first two cases,
by using arguments similar to those in the proof of Lemma~\ref{LASparallel}.
The last case obviously completes gathering.

\noindent
({\bf Case PN})
Suppose that $C(0)$ is in class PN,
which implies that $S(C(0)) \in \{(R,R), (R,W)\}$.
Since a robot $r_i$ at $R$ moves in the negative direction
along its $x$-axis and thus decreases its $x$-coordinate
(without changing its $y$-coordinate)
in its local coordinate system,
$\cal E$ eventually reaches a configuration $C(f)$ in class NN
for the first time at $f$.

If both robots are settled at $f$, 
the case is reduced to Case NN.
If $r_1$ is settled at $f$,
the case is also reduced to Case NN, as follows:
$C(0)$ is in class PN,
$S(C(0)) = $(R, R)$ or (R, W)$,
only $r_0$ is activated at time $0$,
and $r_1$ is activated at time $1$
while $r_0$ is still moving.
Finally, if $r_0$ is settled at $f$,
consider the time $b_1(f)$ 
at which $r_1$ is activated next time after $f$.
It is easy to observe that $C(b_1(f))$ is in NN
and $S(C(b_1(f))) \in \{ (W,A), (R,A)\}$.
Since $r_1$ is settled at $b_1(f)$, as above,
the case is reduced to Case NN.

\noindent
({\bf Case NP})
Suppose that $C(0)$ is in class NP,
which implies that $S(C(0)) \in \{(A,A), (W, A)\}$.
If $S(C(0)) =(W,A)$, then obviously the gathering eventually 
completes.
By similar arguments to those used to show Case PN
and Lemma~\ref{LSSgeneral},
the case is reduced to Case NN, or else
gathering completes.

\noindent
({\bf Case PP})
Suppose that $C(0)$ is in class PP,
which implies that $S(C(0)) = (A,R)$.
Applying arguments similar to Case PN
and the proof of Lemma~\ref{LSSgeneral},
the case is reduced to Case PN,
unless gathering completes.
\hfill $\Box$
\end{pf}

By Lemmas~\ref{LASparallel} and \ref{LASgeneral},
we have the following theorem.

\begin{thm}
\label{TASsufficiency}
For any $0 \leq \phi < \pi/2$,
Algorithm $A_{SS}^{\phi}$ for two oblivious 
robots using $\phi$-static compasses solves the
gathering problem, under the asynchronous model.
\end{thm}

\begin{remark}
At the end of Section~\ref{Ssemisynch},
we modified $A_{SS}^{\phi}$ by replacing the action Gathered into Terminate, 
and showed that the modified $A_{SS}^{\phi}$ is 
a gathering algorithm for semi-synchronous robots
with the termination condition.
The modified $A_{SS}^{\phi}$ however is not a correct gathering algorithm
for asynchronous robots,
as the following counter-example shows.
Let $C(0) = ((0,0),(0,-1)$ and suppose that 
the unit distances of $Z$ and $Z_{(i,t)}$ are the same,
i.e., $sc_i(t) = 1$ for all $ i \in \{0,1 \}$ and $t \in  \Natural$
and that the compasses have no deviation, i.e., $\phi_0 = \phi_1 = 0$.
Then $S(C(0)) = (R,A)$.
Consider the following scenario:
\begin{description}
\item[Time 0:]
$r_0$ and $r_1$ are activated,
where $S(C(0)) = (R,A)$.
\item[Time Interval (0,1):]
$r_1$ moves and reaches $(0,0)$,
but $r_0$ does not move.
\item[Time 1:]
$r_1$ is activated,
where $S(C(0)) = (G,G)$.
Then $r_1$ halts.
\item[Time Interval (1,2):]
$r_0$ moves and reaches $(-1,0)$.
\item[Time 2:]
$r_0$ is activated,
where $S(C(2)) = (W,A)$.
Since $r_1$ has terminated,
neither robot can move.
\end{description}
This shows that the modified $A_{SS}^{\phi}$ 
is not a correct gathering algorithm for asynchronous robots.
\end{remark}

\subsection{Asynchronous Robots with Dynamic Compasses}
\label{SSasynchdynamic}

We present a gathering algorithm $A_{AD}^{\phi}$
for two oblivious asynchronous robots using dynamic compasses,
and show its correctness, provided $0 \leq \phi < \pi/6$.\footnote{
AD of $A_{AD}^{\phi}$ stands for Asynchronous robots with Dynamic compasses.}

\begin{description}
\item[Algorithm $A_{AD}^{\phi}(\mathbf{p})$]~~

\begin{description}
\item[G(athered):]
If $\mathbf{p} = \mathbf{0}$ 
then $A_{AD}^{\phi}(\mathbf{p}) = \mathbf{0}$.
\item[A(pproach):]
If $2\pi/3 + \phi  \leq  \argum(\mathbf{p}) < 3\pi/2$
then $A_{AD}^{\phi}(\mathbf{p}) = \mathbf{p}$.
\item[W(ait):]
If $- \pi/2 (= 3\pi/2)\leq \argum(\mathbf{p}) \leq \pi/3 - \phi$ 
then $A_{AD}^{\phi}(\mathbf{p}) = \mathbf{0}$.
\item[R(otate):]
If $\pi/3 - \phi < \argum(\mathbf{p}) < 2\pi/3 + \phi$
then $A_{AD}^{\phi}(\mathbf{p}) = \rho_{\frac{2\pi}{3} + 2\phi}(\mathbf{p})$. 
\end{description}
\end{description}

\begin{figure}
\begin{center}
\includegraphics[keepaspectratio,width=70mm]{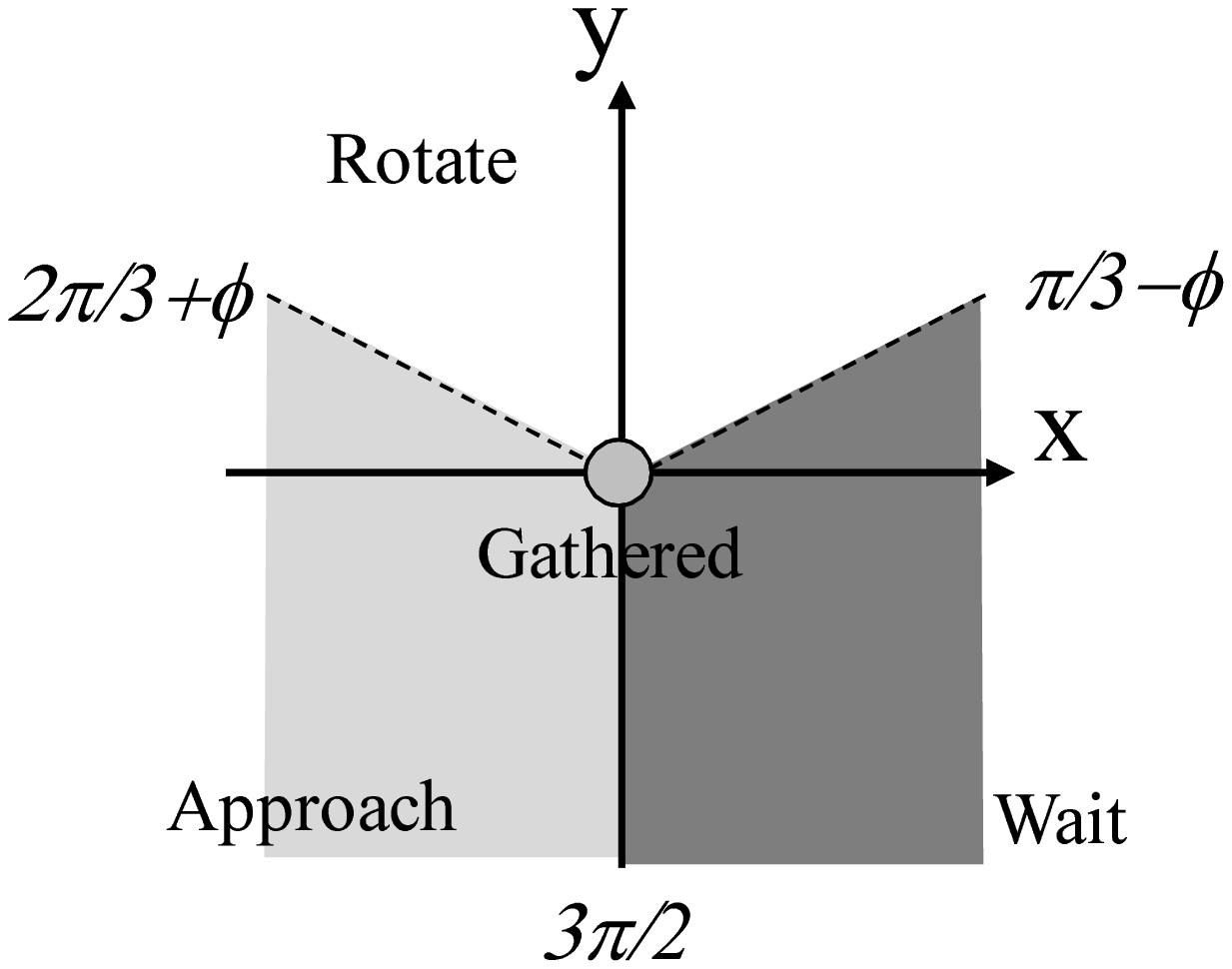}
\caption{An illsutration of Algorithm $A_{AD}^{\phi}$.}
\label{fig:ADcoloredalg}
\end{center}
\end{figure}

\begin{figure}
\begin{center}
\includegraphics[keepaspectratio,width=40mm]{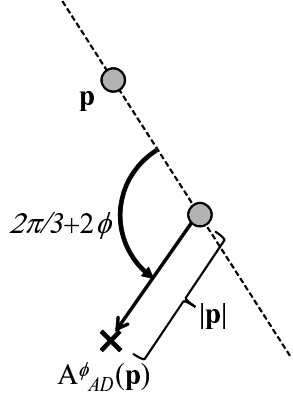}
\caption{The move of a robot in state Rotate who looks the other 
robot at $\mathbf{p}$ in its local coordinate system under 
Algorithm $A_{AD}^{\phi}$.}
\label{fig:ADrotation}
\end{center}
\end{figure}

Figs.~\ref{fig:ADcoloredalg} and \ref{fig:ADrotation} 
illustrate Algorithm $A_{AD}^{\phi}$
and the move of a robot in state $R$ (Rotate) 
who looks the other robot at $\mathbf{p}$ in its local coordinate system 
under Algorithm $A_{AD}^{\phi}$, respectively.    
We show the correctness of $A_{AD}^{\phi}$.

\begin{thm}
\label{TADsufficiency}
For any $0 \leq \phi < \pi/6$,
Algorithm $A_{AD}^{\phi}$ for two oblivious 
robots using $\phi$-dynamic compasses solves the
gathering problem, under the asynchronous model.
\end{thm}

\begin{pf}
Consider any configuration $C(0)$ and any execution 
${\cal E} = C(0), C(1), \ldots$ of $A_{AD}^{\phi}$
with initial configuration $C(0)$.
For any $t \in \Natural$,
let $C(t) = (\mathbf{r}_0(t), \mathbf{r}_1(t))$ and
$\mathbf{r}_i(t) = (x_i(t), y_i(t))$.
By the definition of $A_{AD}^{\phi}$, 
we have $S(C(t)) \in \{ (G,G), (A,W), (W,A), (A,R), (R,A), (W,R), (R,W) \}$
for any $t \in \Natural$.
In the following, 
we show that ${\cal E}$ is a gathering execution.

A configuration $C(t)$ such that $S(C(t)) = (G,G)$ is said to be
{\em pseudo gathered} if $S(C(t')) \not= (G,G)$ for some $t' > t$,
or equivalently, if a robot is not settled at $t$.
Unlike semi-synchronous robots' execution,
${\cal E}$ (of $A_{AD}^{\phi}$) may reach
a pseudo gathered configuration.

Suppose that $C(t)$ is a pseudo gathered configuration
and a robot, say $r_0$, is not settled.
Since the state of $r_1$ is $G$ (i.e., stay motionless),
the execution can reach the same configuration $C(t+1)$ 
even if $r_1$ is not activated at $t$.
Formally, if $C(t)$ is a pseudo gathered configuration,
then ${\cal E}' = C(0), C(1), \ldots, C(t-1), C(t+1), C(t+2), \ldots$
is also an execution of $A_{AD}^{\phi}$.

The proof is by contradiction:
We assume that ${\cal E}$ is not a gathering execution
and derive a contradiction.
If ${\cal E}$ is not a gathering execution,
then there is an execution ${\cal E}'$ such that
it is not a gathering execution and does not contain 
a pseudo gathered configuration.
Without loss of generality, we also assume that pseudo gathering
execution never appear in ${\cal E}$. 

If $y_0(0) = y_1(0)$, 
since $0 \leq \phi < \pi/6$,
$C(0)$ is stable and $S(C(0)) \in \{ (A,W), (W,A), (G,G)\}$
by the definition of $A_{AD}^{\phi}$.
Since the case where $S(C(0)) = (G,G)$ is trivial,
let us assume, without loss of generality, that $S(C(0)) = (A,W)$.
Then obviously, $r_0$ always move toward $r_1$ 
by the definition of $A_{AD}^{\phi}$ 
and the gathering eventually completes.
We thus assume $y_0(0) < y_1(0)$ without loss of generality.

To show the correctness of $A_{SD}^{\phi}$ 
in Subsection~\ref{SSsemisynchdinamic},
we observed that a robot at $R$ rotates the line segment
connecting the current robots' positions clockwise
until the state pair becomes either $(W,A)$ or $(A,W)$.
The scenario of the correctness proof of $A_{AD}^{\phi}$ is similar.
Define $\alpha(t) = \argum(\mathbf{r}_1(t) - \mathbf{r}_0(t))$,
provided that $y_0(t) < y_1(t)$.

For the time being,
we assume (1) $y_0(t) < y_1(t)$ for any $t \in \Natural$
(and hence $0 < \alpha(t) < \pi$), and
(2) $0 < \alpha(t+1) \leq \alpha(t)$.
The verification of their correctness is the core of 
the proof and will be given later. 

Obviously $\alpha(t)$ converges to an angle $\alpha \geq 0$
(under the above two assumptions).
Indeed, $\alpha = 0$; that is, $\alpha(t)$ converges to 0.
To observe this, let us assume that $\alpha > 0$.
Then we can derive a contradiction,
by an argument which is identical to 
the last three paragraphs of 
the proof of Theorem~\ref{TSDsufficiency}.

When $\alpha(t) \approx 0$,  
by the definition of $A_{AD}^{\phi}$,
$C(t)$ is stable\footnote{Recall that a configuration $C$ is
said to be stable if $S(C)$ is determined uniquely, regardless
of the current local coordinate systems $Z_{(i,t)}$.} 
and $S(C(t)) = (W,A)$.
We now show that the gathering eventually completes from such $C(t)$, 
which contradicts the assumption that 
${\cal E}$ is not a gathering execution.

Suppose that $\cal E$ eventually reaches
a configuration $C(t)$ such that $\alpha(t) \approx 0$.
Since $\alpha(t) \approx 0$, 
$C(t)$ is stable and $S(C(t)) = (W,A)$.
Moreover, $C(t')$ is stable and $S(C(t')) = (W,A)$
for all $t' \geq t$.
Let $f = \max \{ b_0(t), b_1(t)\}$.
Then, $r_0$ is settled after (and including) time $f$.
By definition, $\cal E$ eventually reaches $(G,G)$.

Now we return to the verification of 
the two assumptions mentioned above. 
That is, we prove
(1) $y_0(t) < y_1(t)$ and
(2) $0 < \alpha(t+1) \leq \alpha(t)$,
for any $t \in \Natural$. 

To this end, we still need a few more concepts.
Let $s_i(t)$ be the state of robot $r_i$ at time $t$.
That is, letting $S(t) = S(C(t))$, $S(t) = (s_0(t), s_1(t))$.
Since a robot, say $r_0$, may not be settled at $t$,
$s_0(t)$ may not coincide with the action 
$s_0^*(t) (= s_0(a_0(t)))$ that $r_0$ is engaging at $t$.
(For consistency, we assume that $s_i^*(t) = W$ 
if robot $r_i$ has never been activated yet.)
Let $S^*(t) = (s_0^*(t), s_1^*(t))$ = $(s_0(a_0(t)), s_1(a_1(t)))$.

Suppose that $y_0(t) < y_1(t)$.
We partition the working space $\Real^2$ of the robots 
into two half planes delimited by the line $L$ connecting their positions.
Recall that $\alpha(t)$ is the angle that $L$ forms with the $x$-axis of $Z$.
We assume that both half planes contain $L$ as a part,
and denote by  $\Gamma_0(t)$ (resp. $\Gamma_1(t)$)
the left-hand (resp. right-hand) side half plane of $L$.
Robot $r_i$ may or may not be activated at time $t$.
However, if $r_i$ is activated,
it calculates and moves toward the next position,
the coordinates of which are expressed  by $\mathbf{d}_i(t)$ in $Z$.\footnote{
The coordinates of the next position in $Z_{(i,t)}$ is $A_{AD}^{\phi}(\mathbf{p})$, 
where $\mathbf{p}$ represents the coordinates of the other robot in $Z_{(i,t)}$.}

As mentioned, 
we may assume $y_0(0) < y_1(0)$ without loss of generality.
We then prove the following four statements:
For any $t \geq 1$, 
\begin{enumerate}
\item
$y_0(t) < y_1(t)$,
\item
$0 < \alpha(t) \leq \alpha(t-1)$,
\item
$\mathbf{d}_i(t) \in \Gamma_i(t)$ for $i \in \{0, 1\}$, and
\item
$S^*(t) \not= (A,A)$.
\end{enumerate}

Recall that we assume that $\cal E$ is not a gathering execution 
and does not contain a pseudo gathered configuration.
The proof is by induction on $t$.
Since the base case is obvious, 
let us concentrate on the induction step.

(A) First we show $y_0(t) < y_1(t)$.
In the proof, 
we implicitly use the fact that $R$ always decreases 
the robot's $y$-coordinate.
Assume that $y_0(t) \geq y_1(t)$ to derive a contradiction.
Assume first that $r_0$ is activated at $t-1$ ($r_1$ may 
or may not be activated at $t-1$).
Let $v = a_1(t-1) \leq t-1$.
Since $y_0(v) < y_1(v)$, $s_1(v) \in \{ A, W\}$.
If $s_1(v) = W$, then $y_1(t) = y_1(t-1)$.
Since $r_0$ is activated at $t-1$,
$y_0(t) \leq y_1(t-1)$ and 
the equality holds only if $s_0(t-1) = A$.
If $s_0(t-1) = A$, and $y_0(t) = y_1(t-1) = y_1(t)$,
then $S(t) = (G,G)$ (since $y_0(t-1) < y_1(t-1)$),
a contradiction.
If $s_1(v) = A$, 
since $S^*(u) \not= (A,A)$ for all $v \leq u \leq t-1$,
$s_0^*(u) \in \{R, W\}$ for all $v \leq u \leq t-1$,
which implies that $y_0(t) \leq y_0(v) \leq y_1(t)$. If 
$y_0(t) = y_0(v) = y_1(t)$, then we can conclude $S(t) = (G, G)$, 
a contradiction. If $y_0(t) < y_0(v)$ or $y_0(v) < y_1(t)$, 
it directly implies $y_0(t) < y_1(t)$, a contradiction.

Assume next that $r_1$ is activated at $t-1$.
Let $v = a_0(t-1) \leq t-1$.
If $s_0(v) = R$, then $y_0(t) <  y_0(t-1) \leq y_1(t)$,
a contradiction.
If $s_0(v) = W$, then $y_0(v) = y_0(t) = y_0(t-1) < y_1(t-1)$, 
a contradiction is derived, since $s_1(t-1) \in \{A, W\}$ and thus 
$y_0(t) =  y_1(t)$ implies $S(t) = (G,G)$.
If  $s_0(v) = A$,
since $S^*(u) \not= (A,A)$ for all $v \leq u \leq t-1$,
$s_1(u) \not= A$, which implies that $s_1(u)$ is 
always $W$, a contradiction, since $y_0(t) \leq y_1(v) = y_1(t)$
and $y_0(t) =  y_1(t)$ implies $S(t) = (G,G)$.

(B) Second we show $0 < \alpha(t) \leq \alpha(t-1)$.
If both of the robots are settled at $t-1$,
then the claim is obvious, 
since $\mathbf{d}_i(t-1) \in \Gamma_i(t-1)$ for $i \in \{0, 1\}$,
$y_0(t) < y_1(t)$, $0 < \alpha(t) \leq \alpha(t-1)$, and 
$S(t) \not= (G,G)$.

If robot $r_0$ is not settled at $t-1$,
then $\mathbf{d}_0(a_0(t-1)) \in \Gamma_0(a_0(t-1))$ and
$\alpha(t-1) \leq \alpha(a_0(t-1))$,
which implies that $\mathbf{d}_0(a_0(t-1) \in \Gamma_0(t-1)$.
Then by the same argument as above, 
$0  < \alpha(t) \leq \alpha(t-1)$, since $y_0(t) < y_1(t)$.
The case in which $r_1$ is not settled at $t-1$ is symmetrical.

(C) Third we show $\mathbf{d}_i(t) \in \Gamma_i(t)$ 
for $i \in \{0, 1\}$.
Since we showed $y_0(t) < y_1(t)$ in (A) and 
$0 \leq \alpha(t) < \alpha(t-1)$ in (B),
the claim is obvious by the definition of $A_{AD}^{\phi}$.

(D) Finally we show $S^*(t) \not= (A,A)$.
There are two cases to be considered.
Assume first that $r_0$ is activated at $t$.
Since $s_0(t) = A$,
$2\pi/3 \leq \alpha(t) < \pi$
(because $y_0(t) <  y_1(t)$).
Let $v = a_1(t) \leq t-1$.
Since $s_1(v) = A$,
$0 < \alpha(v) \leq \pi/2 + \phi$.
Since $\pi/2 + \phi < 2\pi/3$
(because $\phi < \pi/6$),
a contradiction is derived, 
since $\alpha(t) \leq \alpha(v)$. 

Next assume that $r_1$ is activated at $t$.
Let $v = a_0(t) \leq t-1$.
Since $S^*(u) \not= (A,A)$ for any $v \leq u \leq t-1$,
$s_0(v) = A$,
and $s_1(u) = W$ 
for any $v \leq u \leq t-1$
(since $r_1$ can take either $A$ or $W$),
a contradiction.

\hfill $\Box$
\end{pf}

\section{Concluding Remarks}
\label{Sconc}

This paper investigates the gathering problem for two oblivious 
anonymous mobile robots under disagreement of local coordinate systems. 
To discuss the magnitude of consistency between the local coordinate systems, 
we assumed that each robot is equipped with an unreliable compass, 
the bearings of which may deviate from an absolute reference direction, 
and that the local coordinate system of each robot is determined by its compass.
We considered four classes of robot systems, which are specified by 
the combination of synchrony assumption (semi-synchronous/asynchronous robots)
and compass models (static/dynamic), 
and established the maximum deviation $\phi$
allowing an algorithm to solve the gathering problem for each class:
$\phi < \pi/2$ for semi-synchronous and asynchronous robots with static compasses, 
$\phi < \pi/4$ for semi-synchronous robots with dynamic compasses, 
and $\phi < \pi/6$ for asynchronous robots with dynamic-compasses. 
Except for asynchronous robots with dynamic compasses, 
these sufficient conditions are also necessary.
As for a necessary condition on $\phi$ 
for asynchronous robots with dynamic-compasses
we could show that $\phi < \pi/6$ is necessary for almost all cases,
and thus conjecture it and would like to leave it as a challenging future work.
The results are summarized in Table~\ref{table:gather}.

\begin{conjecture}
Condition $\phi < \pi/6$ is necessary for asynchronous oblivious
robots with dynamic-compasses to have a gathering algorithm.
\end{conjecture}

Remarks 1 and 2 emphasize that the modified $A_{SS}^{\phi}$ is 
a gathering algorithm for semi-synchronous robots 
with the termination property, but is not for asynchronous robots.
An interesting question is hence to ask if there is
a gathering algorithm for asynchronous oblivious robots 
with the termination property.
The gathering process with termination property could be viewed 
as a process of obtaining a point that the robots will gather 
as their common knowledge,
and common knowledge is in general impossible to obtain 
under asynchronous setting.
The plausible answer is thus ``NO,''
and we would like to conjecture it.
However, in order to complete a proof,
we first need to deeply understand 
why gathering with termination is possible for semi-synchronous robots,
despite that they share some asynchronous nature with asynchronous robots.

As a final note, in \cite{YS09}, the authors show that there is no 
gathering algorithm for oblivious robots under the semi-synchronous
model even if the symmetricity of the initial configuration is 1
(i.e., even if the deviation of their local coordinate systems is 
less than $\pi$). We would like to note that this fact does not 
contradict to Theorem \ref{TSSsufficency} since Algorithm $A_{SS}^{\phi}$
relies on the existence of upper bound $\phi$.

\begin{table}
  \centering
  \caption{The summary of results about two oblivious-robot gathering 
with unreliable compasses.}
  \label{table:gather}
  \medskip
\begin{tabular}{|ccc|c|c|l}

\cline{4-5} 
\multicolumn{3}{c}{} & \multicolumn{2}{|c|}{Compass} \\ \cline{4-5}  
\multicolumn{3}{c|}{} & Static &  Dynamic  \\ \cline{1-5}
\multirow{4}{*}{Timing model}&
\multicolumn{1}{|c|}{\multirow{2}{*}{\emph{S.Synch}}}  &
\multicolumn{1}{|c|}{Possible} & $\phi < \pi/2$  
(\textsl{Sec.~\ref{SSsemisynchstatic}})
& $\phi <\pi/4$ (\textsl{Sec.~\ref{SSsemisynchdinamic}}) & \\
\cline{3-5}

& \multicolumn{1}{|c|}{} &
\multicolumn{1}{|c|}{Impossible} & $ \phi \geq \pi/2$ (\textsl{~\cite{P05,SY99}}) & $ \phi \geq \pi/4$ (\textsl{Sec.~\ref{SSsemisynchdinamic}}) & \\
\cline{2-5}

& \multicolumn{1}{|c|}{\multirow{2}{*}{\emph{Asynchronous}}} &
\multicolumn{1}{|c|} {Possible} & $\phi < \pi/2$  
(\textsl{Sec.~\ref{SSasynchstatic}}) 
& $\phi < \pi/6$  (\textsl{Sec.~\ref{SSasynchdynamic}}) & \\
\cline{3-5}

& \multicolumn{1}{|c|}{} &
\multicolumn{1}{|c|}{Impossible} & $\phi \geq \pi/2$  (\textsl{deduction}) & $\phi \geq \pi/4$  (\textsl{deduction}) & \\
\cline{1-5}
\end{tabular}
  \end{table}

\end{document}